\documentclass[12pt]{article}

\usepackage {graphicx}
\usepackage{amssymb}
\usepackage{amsmath}
\usepackage{amsbsy}
\usepackage{setspace}
\usepackage[square, sort, comma, numbers]{natbib}
\usepackage{bm}
\usepackage{textcomp}
\usepackage{color}

\usepackage[vlined]{algorithm2e}
\usepackage{amsthm}
\usepackage{enumitem}
\usepackage{longtable,threeparttablex,booktabs}
\usepackage{hyperref}
\usepackage{mathtools}
\usepackage{lscape}
\usepackage{subfigure}
\usepackage{multirow}

\setcitestyle{square}

\newtheorem{theorem}{Theorem}
\newtheorem{corollary}{Corollary}
\newtheorem{proposition}{Proposition}
\newtheorem{lemma}{Lemma}
\newtheorem{example}{Example}

\newtheorem{remark}{Remark}
\newtheorem{definition}{Definition}

\newtheorem{rem}{Remark}[section]

\def\text#1{\mbox{\rm #1}}
\def\overset#1#2{\stackrel{#1}{#2} }

\def\underwiggle 1{
\ifmmode\setbox\TempBox=\hbox{$ 1$}\else\setbox\TempBox=\hbox{
1}\fi \setbox\TempBoxA=\hbox to \wd\TempBox{\hss\char'176\hss}
\rlap{\copy\TempBox}\smash{\lower9pt\hbox{\copy\TempBoxA}} }

\newcommand{\beq}{\begin{equation}}
\newcommand{\eeq}{\end{equation}}
\newcommand{\beas}{\begin{eqnarray*}}
\newcommand{\eeas}{\end{eqnarray*}}
\newcommand{\bea}{\begin{eqnarray}}
\newcommand{\eea}{\end{eqnarray}}
\newcommand{\bei}{\begin{itemize}}
\newcommand{\eei}{\end{itemize}}
\newcommand{\ben}{\begin{enumerate}}
\newcommand{\een}{\end{enumerate}}
\newcommand{\bet}{\begin{theorem}}
\newcommand{\eet}{\end{theorem}}
\newcommand{\bel}{\begin{lemma}}
\newcommand{\eel}{\end{lemma}}
\newcommand{\bep}{\begin{proposition}}
\newcommand{\eep}{\end{proposition}}
\newcommand{\bed}{\begin{definition}}
\newcommand{\eed}{\end{definition}}
\newcommand{\bec}{\begin{corollary}}
\newcommand{\eec}{\end{corollary}}
\newcommand{\bex}{\begin{example}}
\newcommand{\eex}{\end{example}}

\newcommand{\argmin}{\mathop{\rm arg\min}}
\newcommand{\argmax}{\mathop{\rm arg\max}}

\def\limsup{\mathop{\overline{\rm lim}}}

\begin{document}
\title{Optimal False Discovery Rate Control for Large Scale Multiple Testing with Auxiliary Information}

\author{Hongyuan Cao$^1$, Jun Chen$^2$ and Xianyang Zhang$^3$}

\footnotetext[1]{~ Department of Statistics, Florida State University, Tallahassee, FL, U.S.A.}
\footnotetext[2]{~ Division of Biomedical Statistics, Mayo Clinic, Rochester, MN, U.S.A.}
%\footnotetext[2]{~ Department of Statistics and Applied Probability, National University of Singapore, 117546, Singapore}
% \newline \indent
%\ \ \ \ The research of Tony Cai was supported in part by NSF Grant
%DMS-0604954.}
\footnotetext[3]{~Department of Statistics, Texas A\&M University, College Park, U.S.A. The authors are ordered alphabetically. Correspondence should be addressed to Xianyang Zhang (zhangxiany@stat.tamu.edu)}
\date{}
\maketitle

\begin{abstract}
Large-scale multiple testing is a fundamental problem in high dimensional statistical inference. It is increasingly common that various types of auxiliary information, reflecting the  structural relationship among the hypotheses, are available. Exploiting such auxiliary information can boost statistical power.
To this end, we propose a framework based on a two-group mixture model
with varying probabilities of being null for different hypotheses \textit{a priori}, where a shape-constrained relationship is imposed between the auxiliary information and the prior probabilities of being null.  An optimal rejection rule is designed to maximize the expected number of true positives when average false discovery rate is controlled.  Focusing on the ordered structure, we develop a robust EM algorithm to estimate the prior probabilities of being null and the distribution of $p$-values under the alternative hypothesis simultaneously. We show that the proposed method has better power than state-of-the-art competitors while controlling the false discovery rate, both empirically and theoretically. Extensive simulations demonstrate the advantage of the proposed method.
Datasets from genome-wide association studies are used to illustrate the new methodology.

\end{abstract}

\noindent \textbf{Keywords: \/}
EM algorithm, False discovery rate, Isotonic regression, Local false discovery
rate, Multiple testing, Pool-Adjacent-Violators algorithm.
\thispagestyle{empty}

\section{Introduction}\label{intro.sec}
\setcounter{page}{1}
%Large scale multiple testing problems have been extensively explored and are now well understood in both theory and practice \cite{bh95, genoves2004, sts2004, sc07}. With technological advancement, various types of auxiliary information are available, particularly for genomic data. For instance, in metagenomic data analysis, a phylogenetic tree provides important evolutionary relationships among various biological species  based upon similarities and differences in their physical or genetic characteristics \cite{woese2002}. In epigenomic studies, genome locations provide important information about the prior correlation structure of the epigenetic markers \cite{bumphunting2012}. In integrative genomic analysis, multiple ``omics" are combined to decipher the disease etiology and association \cite{integrative2014}. Incorporating such rich and informative auxiliary information can boost statistical power.  %Auxiliary data contains useful information that can be integrated into multiple testing procedures.
%one set of ``omics" data could provide important gene regulation information for more powerful analysis of the other set. Ignoring such auxiliary information can cause loss of statistical power.
%To incorporate prior or structural information,
Large scale multiple testing refers to simultaneously testing of many hypotheses. Given a pre-specified significance level, family-wise error rate (FWER) controls the probability of making one or more false rejections, which can be unduly conservative in many applications. The false discovery rate (FDR) controls the expected value of the false discovery proportion, which is defined as the ratio of the number of false rejections divided by the number of total rejections. Benjamini and Hochberg (BH)
\cite{bh95} proposed a FDR control procedure that sets adaptive thresholds for the $p$-values.
 %
%Multiple testing refers to simultaneous testing of more than one hypothesis. Given a (large) set of hypotheses, multiple testing approach often concerns about deciding which hypotheses to reject while guaranteeing some notion of control on the number of false rejections. Both family-wise error rate (FWER) and false discovery rate (FDR) have been proposed for controlling false rejections. Compared to FWER control procedures, which control the  probability of making one or more false rejections, FDR control procedures are substantially more powerful by tolerating some level of false rejections in the rejected hypotheses. Thus FDR control is a popular multiple testing approach for producing a relatively reliable and replicable list of discoveries with reduced experimental cost. Given $m$ hypotheses and a rejection rule, suppose a total of $R$ hypotheses are rejected, among which $V$ are false rejections. The FDR is defined as $E[V/\max\{R,1\}]$.
%
%Perhaps the most celebrated multiple testing procedure of the modern era is the Benjamini-Hochberg (BH) procedure \cite{bh95}, which intends to control the FDR by setting an adaptive threshold for the $p$-values.
It turns out that the actual FDR level of the BH procedure is the multiplication of the proportion of null hypotheses and the pre-specified significance level. Therefore, the BH procedure can be overly conservative when the proportion of null hypotheses is far from one. To address this issue, \cite{storey2003} proposed a two-stage procedure (ST), which first estimates the proportion of null hypotheses %based on the percentage of large p-values
and uses the estimated proportion to adjust the threshold in the BH procedure at the second stage. From an empirical Bayes perspective,  \cite{efron2001} proposed the notion of local FDR (Lfdr) based on the two-group mixture model. \cite{sc07} developed a step-up procedure based on Lfdr and demonstrated its optimality from the compound decision viewpoint. % of compound-decision problem.

The aforementioned methods are based on the premise that the hypotheses are exchangeable. However, in many scientific applications, particularly in genomics,  auxiliary information regarding the pattern of signals is available. For instance, in differential expression analysis of RNA-seq data, which tests for difference in the mean expression of the genes between conditions, the sum of read counts per gene across all samples could be the auxiliary data since it is informative of the statistical power \cite{love2014}.  In differential abundance analysis of microbiome sequencing data, which tests for difference in the mean abundance of the detected bacterial species between conditions,  the genetic divergence among species is important auxiliary information, since closely-related species usually have similar physical characteristics and tend to covary with the condition of interest \cite{xiao2016}.    In genome-wide association studies, the major objective is to test for association between the genetic variants and a phenotype of interest.   The minor allele frequency and the pathogenicity score of the genetic variants, which are informative of the statistical power and the prior null probability, respectively,  are potential auxiliary data, which could be leveraged to improve the statistical power as well as enhance interpretability of the results. 

Accommodating auxiliary information in multiple testing has recently been a very active research area. %  topic of substantial statistical interest.
Many methods have been developed adapting to different types of structure among the hypotheses. The basic idea is to relax the $p$-value thresholds for hypotheses that are more likely to be alternative and tighten the thresholds for the other hypotheses so that the overall FDR level can be controlled.  For example, \cite{genovese2006} proposed to weight the $p$-values with different weights, and then apply the BH procedure to the weighted $p$-values.
\cite{hu2010} developed a group BH procedure by estimating the proportion of null hypotheses for each group separately. \cite{lb17} generalized this idea by using the censored $p$-values (i.e., the $p$-values that are greater than a pre-specified threshold) to adaptively estimate the weights that can be designed to reflect any structure believed to be present. \cite{Ignatiadis2016,Ignatiadis2017} proposed the independent hypothesis weighting (IHW) for multiple testing with covariate information. The idea is to use cross-weighting to achieve finite-sample FDR control. Note that the binning in IHW is only to operationalize the procedure and it can be replaced by the proposed EM algorithm below. 

The above procedures can be viewed to some extent as different variants of the weighted-BH procedure. Another closely related method was proposed in \cite{lf16}, which iteratively estimates the $p$-value threshold using partially masked $p$-values. It can be viewed as a type of Knockoff procedure \cite{BC2015} that uses the symmetry of the null distribution to estimate the false discovery proportion.

Along a separate line, Lfdr-based approaches have been developed to accommodate various forms of auxiliary information. For example, \cite{cai09} considered multiple testing of grouped hypotheses. The authors proposed an optimal data-driven procedure that uniformly improves the pooled and separate analyses. \cite{sunspatial15} developed an Lfdr-based method to incorporate spatial information. \cite{scott2015,tansey2015} proposed EM-type algorithms to estimate the Lfdr by taking into account covariate and spatial information, respectively.

Other related works include \cite{Ferk2008}, which considers the two-group mixture models with side-information. \cite{Dob2017} develops a method for estimating the constrained optimal weights for Bonferroni multiple testing. \cite{BL2018} proposes an FDR-controlling procedure based on the covariate-dependent null probabilities.

In this paper, we develop a new method along the line of research on Lfdr-based approaches by adaptively estimating the prior probabilities of being null in Lfdr that reflect auxiliary information in multiple testing. The proposed Lfdr-based procedure is built on the optimal rejection rule as shown in Section \ref{sec:setup} and thus is expected to be more powerful than the weighted-BH procedure when the underlying two-group mixture model is correctly specified. Compared to existing work on Lfdr-based methods, our contributions are three-fold. (i)
We outline a general framework for incorporating various forms of auxiliary information. This is achieved by allowing the prior probabilities of being null to vary across different hypotheses.
We propose a data-adaptive step-up procedure and show that it provides asymptotic FDR control when relevant consistent estimates are available. (ii) Focusing on the ordered structure,  where auxiliary information generates a ranked list of hypotheses, we develop a new EM-type algorithm \cite{D77} to estimate the prior probabilities of being null and the distribution of $p$-values under the alternative hypothesis simultaneously. Under monotone constraint on the density function of $p$-values under the alternative hypothesis, we utilize the Pool-Adjacent-Violators Algorithm (PAVA) to estimate both the prior probabilities of being null and the density function of $p$-values under the alternative hypothesis (see \cite{G56} for early work on this kind of problems). Due to the efficiency of PAVA, our method is scalable to large datasets arising in genomic studies. (iii) We prove asymptotic FDR control for our procedure and obtain some consistency results for the estimates of the prior probabilities of being null and the alternative density, which is of independent theoretical interest.
Finally, to allow users to conveniently implement our method and reproduce the numerical results reported in Sections \ref{sec:sim}-\ref{sec:data}, we make our code publicly available at \url{https://github.com/jchen1981/OrderShapeEM}.

%https://github.com/jchen1981/OrderShapeEM.

The problem we considered is related but different from the one in \cite{gwct,lb16}, where the authors seek the largest cutoff $k$ so that one rejects the first $k$ hypotheses while accepts the remaining ones. So their method always rejects an initial block of hypotheses. In contrast, our procedure allows researchers to reject the $k$th hypothesis but accept the $k-1$th hypothesis in the ranked list. In other words, we do not follow the order restriction strictly. Such flexibility could result in a substantial power increase when the order information is not very strong or even weak, as observed in our numerical studies. Also see the discussions on monotonicity in Section 1.1 of \cite{scott2015}.

To account for the potential mistakes in the ranked list or to improve power by incorporating external covariates, alternative methods have been proposed in the literature. For example, \cite{lynch2017} extends the fixed sequence method to allow more than one acceptance before stopping. \cite{lei2018}
modifies AdaPT in \cite{lf16} by giving analysts the power to enforce the ordered constraint on the final rejection set. Though aiming for addressing a similar issue, our method is motivated from the empirical Bayes perspective, and it is built on the two-group mixture model that allows the prior probabilities of being null to vary across different hypotheses. The implementation and theoretical analysis of our method are also quite different from those in \cite{lei2018,lynch2017}.

Finally, it is also worth highlighting the difference with respect to the recent work \cite{deb2019} which is indeed closely related to ours. First of all, our Theorem 3.3 concerns about the two-group mixture models with decreasing alternative density, while Theorem 3.1 in \cite{deb2019} focuses on a mixture of Gaussians. We generalize the arguments in \cite{van2000} by considering a transformed class of functions to relax the boundedness assumption on the class of decreasing densities. A careful inspection of the proof of Theorem 3.3 reveals that the techniques we develop are quite different from those in \cite{deb2019}. Second, we provide a more detailed empirical and theoretical analysis of the FDR-controlling procedure. In particular, we prove that the step-up procedure based on our Lfdr estimates asymptotically controls the FDR and provide the corresponding power analysis. We also conduct extensive simulation studies to evaluate the finite sample performance of the proposed Lfdr-based procedure.

The rest of the paper proceeds as follows. Section \ref{sec:test} proposes a general multiple testing procedure
that incorporates auxiliary information to improve statistical
power, and establishes its asymptotic FDR control property. In Section \ref{sec:iso}, we introduce a new EM-type algorithm to
estimate the unknowns and study the theoretical properties
of the estimators.
%We describe a general rejection rule and make a connection %with the accumulation tests in Section
%\ref{sec:pval}. %We study the
%Asymptotic power of the proposed method is examined in %Section \ref{sec:power}.
We discuss two extensions in Section \ref{sec:extend}. Section
\ref{sec:sim} and Section \ref{sec:data} are devoted respectively to
simulation studies and data analysis. % to investigate the finite
%sample performance of the proposed method.
We conclude the paper in
Section \ref{sec:summary}. All the proofs of the main theorems and
technical lemmas are collected in the Appendix.

\section{Covariate-adjusted multiple testing}\label{sec:test}
In this section, we describe a covariate-adjusted multiple testing
procedure based on Lfdr.

\subsection{Optimal rejection rule}\label{sec:setup}
Consider simultaneous testing of $m$ hypotheses $H_i$ for $i=1,
\ldots,m$ based on $m$ $p$-values $x_1,\dots,x_m$, where $x_i$ is the
$p$-value corresponding to the $i$th hypothesis $H_i.$ Let $\theta_i,
i =1, \ldots, m$ indicate the underlying truth of the $i$th
hypothesis. In other words, $\theta_i = 1$ if $H_i$ is non-null/alternative and
$\theta_i=0$ if $H_i$ is null. We allow the probability
that $\theta_i=0$ to vary across $i$. In this way, auxiliary information
can be incorporated through
\begin{equation} \label{prior-prob}
P(\theta_i = 0) = \pi_{0i}, \quad i = 1, \ldots, m.
\end{equation}
Consider the two-group model for the $p$-values (see e.g., \cite{Efron2008} and Chapter 2 of \cite{Efron2012}):
\begin{equation}\label{mix-model}
x_i \mid \theta_i\sim (1-\theta_i)f_0 + \theta_i f_1, \quad i =1,
\ldots, m,
\end{equation}
where $f_0$ is the density function of the $p$-values under the null hypothesis and $f_1$
is the density function of the $p$-values under the alternative hypothesis. The marginal probability
density function of $x_i$ is equal to
\begin{align}\label{eq-iden}
f^i(x) = \pi_{0i}f_0(x) +
(1-\pi_{0i})f_1(x).
\end{align}

We briefly discuss the identifiability of the above model. Suppose $f_0$ is known and bounded away from zero and infinity. Consider the following class of functions:
\begin{align*}
\mathbf{F}_m=&\big\{\tilde{\mathbf{f}}=(\tilde{f}^1,\dots,\tilde{f}^m) \text{ with } \tilde{f}^i=\tilde{\pi}_if_0+(1-\tilde{\pi}_i)\tilde{f}_1:\min_{x\in [0,1]} \tilde{f}_1(x)=0,
\\ &0\leq \tilde{\pi}_i\leq 1,\min_i \tilde{\pi}_i<1\}.
\end{align*}
Suppose $\tilde{\mathbf{f}},\breve{\mathbf{f}}\in \mathbf{F}_m$, where the $i$th components of
$\tilde{\mathbf{f}}$ and $\breve{\mathbf{f}}$ are given by $\tilde{f}^i=\tilde{\pi}_if_0+(1-\tilde{\pi}_i)\tilde{f}_1$ and $\breve{f}^i=\breve{\pi}_if_0+(1-\breve{\pi}_i)\breve{f}_1$ respectively. We show that if $\tilde{f}^i(x)=\breve{f}^i(x)$ for all $x$ and $i$, then $\tilde{f}_1(x)=\breve{f}_1(x)$
and $\tilde{\pi}_i=\breve{\pi}_i$ for all $x$ and $i$. Suppose $\tilde{f}_1(x')=0$ for some $x'\in[0,1]$. If $\tilde{\pi}_{i}<\breve{\pi}_{i}$ for some $i$, then we have
\begin{align}\label{iden-1}
0=\frac{\tilde{f}_1(x')}{f_0(x')}=\frac{\breve{\pi}_{i}-\tilde{\pi}_{i}}{1-\tilde{\pi}_{i}}+\frac{(1-\breve{\pi}_{i})\breve{f}_1(x')}{(1-\tilde{\pi}_{i})f_0(x')}>0,
\end{align}
which is a contradiction. Similarly, we get a contradiction when $\tilde{\pi}_{i}>\breve{\pi}_{i}$ for some $i$. Thus we have $\tilde{\pi}_i=\breve{\pi}_i$ for all $i$. As there exists a $i$ such that $1-\tilde{\pi}_i=1-\breve{\pi}_i>0$, it is clear that
$\tilde{f}^i(x)=\breve{f}^i(x)$ implies that $\tilde{f}_1(x)=\breve{f}_1(x)$.

In statistical and scientific applications, the
goal is to separate the alternative cases ($\theta_i=1$) from the null
cases ($\theta_i=0$). This can be formulated as a multiple testing
problem, with solutions represented by a decision rule
$\boldsymbol{\delta} = (\delta_1, \ldots, \delta_m) \in \{0, 1
\}^m.$ It turns out that the optimal decision rule is closely related to
the Lfdr defined as
\begin{equation*}%\label{lfdr}
\mbox{Lfdr}_i(x) := P(\theta_i=0 \mid x_i=x) =
\frac{\pi_{0i}f_0(x)}{\pi_{0i}f_0(x) + (1-\pi_{0i})f_1(x)} =
\frac{\pi_{0i}f_0(x)}{f^i(x)}.
\end{equation*}
In other words, $\mbox{Lfdr}_i(x)$ is the posterior probability that a
case is null given the corresponding $p$-value is equal to $x$. It
combines the auxiliary information ($\pi_{0i}$) and data from the
current experiment. Information across tests is used in forming $f_0(\cdot)$ and
$f_1(\cdot).$ %Consequently, it has good interpretation from the Bayesian perspective.

Optimal decision rule under mixture model has been extensively studied in the literature, see e.g., \cite{tz,lf16,basu2017}. For completeness, we present the derivations below and remark that they follow somewhat directly from existing results. Consider the expected number of false
positives (EFP) and true positives (ETP) of a decision rule.
Suppose that $x_i$ follows the mixture model (\ref{mix-model}) and
we intend to reject the $i$th null hypothesis if $x_i\le c_i.$ The
size and power of the $i$th test are given respectively by
\begin{equation*}
\alpha_i(c_i) = \int_{0}^{c_i}f_0(t)dt \quad \mbox{and} \quad \beta_i(c_i) =
\int_{0}^{c_i}f_1(t)dt.
\end{equation*}
It thus implies that
\begin{equation*}
\mbox{EFP}(\mathbf{c}) = \sum_{i=1}^{m}\pi_{0i}\alpha_i(c_i)   \quad \mbox{and} \quad \mbox{ETP}(\mathbf{c}) = \sum_{i=1}^{m}(1-\pi_{0i})\beta_i(c_i),
  \end{equation*}
where $\mathbf{c} = (c_1, \ldots, c_m).$ We wish to maximize ETP for a given value of the marginal FDR
(mFDR) defined as
\begin{align}\label{mfdr}
\mbox{mFDR}(\mathbf{c})=\frac{\mbox{EFP}(\mathbf{c})}{\mbox{ETP}(\mathbf{c})+\mbox{EFP}(\mathbf{c})},
\end{align}
by an optimum choice of the cutoff value $\mathbf{c}.$ Formally, consider the problem
\begin{align}\label{opt}
\max_{\mathbf{c}}\mbox{ETP}(\mathbf{c}) \quad \text{subject to}\quad \text{mFDR}(\mathbf{c})\leq \alpha.
\end{align}
A standard Lagrange multiplier argument gives the following result which motivates our choice of thresholds.
\begin{proposition}\label{opt-prop}
Assume that $f_1$ is continuously non-increasing, and $f_0$ is continuously non-decreasing and uniformly bounded from above. Further assume that for a pre-specified $\alpha>0,$
\begin{equation}\label{eq-thm2}
\min_{i}\frac{(1-\pi_{0i})f_1(0)}{\pi_{0i}f_0(0)}>\frac{1-\alpha}{\alpha}.
\end{equation}
Then (\ref{opt}) has at least one solution and every solution $(\tilde{c}_1,\dots,\tilde{c}_m)$ satisfies %that
$$\mbox{Lfdr}_i(\tilde{c}_i)=\tilde{\lambda}$$
for some $\tilde{\lambda}$ that is independent of $i.$
\end{proposition}
The proof of Proposition \ref{opt-prop} is similar to that of Theorem 2 in \cite{lf16} and we omit the details. Under the monotone likelihood ratio
assumption \cite{sc07,cao13}:
\begin{equation}\label{mlr}
f_1(x)/f_0(x)\mbox{ is decreasing in }x,
\end{equation}
we obtain that $\mbox{Lfdr}_i(x)$ is monotonically increasing in
$x.$ Therefore, we may reduce our attention to the rejection rule
$\mathbf{I}\{x_i \leq c_i\}$ as
\begin{equation}\label{rejection}
\delta_i = \mathbf{I}\{ \mbox{Lfdr}_i(x_i) \le \lambda\}
\end{equation}
for a constant $\lambda$ to be determined later.

\subsection{Asymptotic FDR control}\label{sec:asym}
To fully understand the proposed method, we gradually
investigate its theoretical properties through several steps,
starting with an oracle procedure which provides key insights into
the problem. %First, we consider an oracle procedure that serves as a
%heuristic device. We a
Assume that $\{\pi_{0i}\}^{m}_{i=1},
f_0(\cdot)$ and $f_1(\cdot)$ are known. The proposed method utilizes
auxiliary information through $\{\pi_{0i}\}^{m}_{i=1}$ and information
from the alternative through $f_1(\cdot)$ in addition to information
from the null, upon which conventional
approaches are based. In view of (\ref{rejection}), the number
of false rejections equals to
$$
V_m(\lambda) = \sum_{i=1}^{m}\mathbf{I} \{\mbox{Lfdr}_i(x_i) \le
\lambda \}(1-\theta_i)
$$
and the total number of rejections is given by
$$
D_{m,0}(\lambda) = \sum_{i=1}^{m}\mathbf{I}\{\mbox{Lfdr}_i(x_i) \le
\lambda \}.
$$

%In this section, we consider an oracle case that serves as a heuristic device
%We now consider an ordered structure on $\pi_{i0}.$ Without loss of generality, we assume that
%\begin{equation}
%\pi_{01}\le \pi_{02} \le \cdots \le \pi_{0m}.
%\end{equation}
%Such ordering information can be obtained from (\ref{proportion}), for example.
%We now consider an oracle procedure that serves as a heuristic device based on local false discovery rate. In this section, we ssume that $\pi_{i0}, i =1, \ldots, m, f_0(\cdot)$ and $f_1(\cdot)$ are known. The proposed method utilizes prior information through $\pi_{i0}, i =1, \ldots, m$ and information from the non-null through $f_1(\cdot).$

Write
$a\vee b=\max\{a,b\}$ and $a\wedge b=\min\{a,b\}$. We aim to find the critical value $\lambda$ in (\ref{rejection})
that controls the FDR, which is defined as
$\text{FDR}_m(\lambda)=E\{V_m(\lambda)/(D_{m,0}(\lambda)\vee 1)\}$ at a
pre-specified significance level $\alpha.$ %We n
Note that
\begin{align}\label{eq-1}
E[V_m(\lambda)]=&\sum_{i=1}^{m}\pi_{0i}P(\mbox{Lfdr}_i(x_i) \le
\lambda
|\theta_i=0)=\sum_{i=1}^{m}E[\mbox{Lfdr}_i(x_i)\mathbf{I}\{\mbox{Lfdr}_i(x_i)
\le \lambda \}].
\end{align}
An estimate of the $\text{FDR}_m(\lambda)$ is given by
\begin{align*}
\mbox{FDR}_m(\lambda)=\frac{\sum_{i=1}^{m}\mbox{Lfdr}_i(x_i)\mathbf{I}\{\mbox{Lfdr}_i(x_i)
\le \lambda \}}{\sum_{i=1}^{m}\mathbf{I}\{\mbox{Lfdr}_i(x_i) \le
\lambda \}}.
\end{align*}
Let $\lambda_m=\sup\{\lambda\in [0,1]: \mbox{FDR}_m(\lambda)\leq \alpha\}$.
Then reject $H_i$ if $\text{Lfdr}_i(x_i)\leq \lambda_m.$ Below we
show that the above (oracle) step-up procedure provides asymptotic
control on the $\text{FDR}$ under the following assumptions.
\begin{enumerate}
\item[(C1)]
Assume that for any $\lambda\in [0,1]$,
\begin{align*}
&\frac{1}{m}\sum_{i=1}^{m}
\mathbf{I}\{\mbox{Lfdr}_i(x_i) \le \lambda \}\rightarrow^p D_0(\lambda),\\
&\frac{1}{m}\sum_{i=1}^{m}\mbox{Lfdr}_i(x_i)\mathbf{I}\{\mbox{Lfdr}_i(x_i)
\le \lambda\}\rightarrow^p D_1(\lambda),
\end{align*}
and
\begin{equation}\label{limit-nonnull}
\frac{1}{m}V_m(\lambda)\rightarrow^p D_1(\lambda),
\end{equation}
where $D_0$ and $D_1$ are both continuous functions over $[0,1]$.

\item [(C2)]
Write $R(\lambda)=D_1(\lambda)/D_0(\lambda)$, where $D_0$ and $D_1$ are defined in (C1). There exists a $\lambda_{\infty} \in (0,1]$ such that $R(\lambda_{\infty})<\alpha.$ %where $\alpha$
%
%\item[(C2)]
%For each $t\in (0, 1],$
%\begin{equation}\label{limit-null}
%D_0(\lambda) := \mbox{lim}_{m\rightarrow \infty} \frac{1}{m}\sum_{i=1}^{m}I\{\mbox{Lfdr}_i(x_i) \le \lambda \mid \theta_i=0\} \pi_{0i} \quad a.s.% = D_0(\lambda)
%\end{equation}
%and
%\begin{equation}\label{limit-nonnull}
%D_1(\lambda):=\mbox{lim}_{m\rightarrow \infty}\frac{ 1}{m}\sum_{i=1}^{m}I\{\mbox{Lfdr}_i(x_i) \le \lambda \mid \theta_i=1 \}(1-\pi_{0i}) \quad a.s., %= D_1(\lambda).
%\end{equation}
%where $D_0$ and $D_1$ are continuous functions.
%\item[(C1)] $P(\theta_i = 0) = \pi_{0i},$ $x_i\mid \theta_i $ are independent, has pdf $f_i = \pi_{0i}f_0 + (1-\pi_{0i})f_1, i=1, \ldots, m,$ and (\ref{monotone}) holds. Suppose that $f_i, i =1, \ldots, m$ are continuous and positive on the real line.
%Assume that $x_i, i =1, \ldots, m$ are independent, (\ref{monotone}) holds and
%\item[(C2)] %In addition, t
%There exist constants $c_1$ and $c_2$, such that $0<c_1 \le \pi_{0i}\le c_2<1, i = 1, \ldots, m.$

%\item[(C3)]
%\begin{equation}\label{monotone}
%f_1(x)/f_0(x) \quad \mbox{is monotonically decreasing in} \quad x.
%\end{equation}

%\item [(C4)]
%There exist a $\lambda_{\infty} \in (0,1],$ such that $D_0(\lambda_{\infty})/\{D_0(\lambda_{\infty}) + D_1(\lambda_{\infty}) \} <\alpha.$ %where $\alpha$
%\item[(C3)] Suppose that $D_0(\lambda):=\lim_{m\rightarrow \infty}\frac{1}{m}\sum_{i=1}^{m}\pi_{0i}F_0(g(a_i\frac{\lambda}{1-\lambda}))$ and \newline
%$D_1(\lambda):=\lim_{m\rightarrow \infty} \frac{1}{m}\sum_{i=1}^{m}(1-\pi_{0i})F_1(g(a_i\frac{\lambda}{1-\lambda}))$ exist.
\end{enumerate}

We remark that (C1) is similar to those for Theorem 4 in \cite{sts2004}. In view of (\ref{eq-1}),
(\ref{limit-nonnull}) follows from the weak law of large numbers.
Note that (C1) allows certain forms of dependence, such as
$m$-dependence, ergodic dependence and certain mixing type
dependence. (C2) ensures the existence of the critical value $\lambda_m$ to asymptotically control the FDR at level $\alpha.$ The following proposition shows
that the oracle step-up procedure provides asymptotic FDR control.

%\begin{eqnarray*}
%\mbox{aFDR}_m& = &\frac{E(V)}{E(R)}\\
%& =& \frac{\sum_{i=1}^{m} \{ \pi_{0i}P(\mbox{Lfdr}_i(x_i) \le \lambda \mid \theta_i = 0) \}}{\sum_{i=1}^{m}\{\pi_{0i}P(\mbox{Lfdr}_i(x_i) \le \lambda \mid \theta_i=0) + (1-\pi_{0i}) P(\mbox{Lfdr}_i(x_i) \le \lambda \mid \theta_i = 1)\}} .
%\end{eqnarray*}
%Denote $g(t) = (f_0/f_1)^{-1}(t)=\inf \{u: f_0(u)/f_1(u)\ge t \},$ $a_i = (1-\pi_{0i})/\pi_{0i},$ and $F_0$ and $F_1$ as cdf of $f_0$ and $f_1$, respectively.
\begin{proposition} \label{fdr-control}
Under conditions (C1)-(C2),
\begin{equation*}
\limsup_{m\rightarrow \infty}\mbox{FDR}_m(\lambda_m)\le \alpha.
\end{equation*}
\end{proposition}
The proof of Proposition \ref{fdr-control} is relegated in the
Appendix. In the following, we mimic the operation of the oracle
procedure and provide an adaptive procedure. In the inference
problems that we are interested in, the $p$-value distribution under the null hypothesis is assumed
to be known (e.g., the uniform distribution on $[0,1]$, or can be obtained from the distributional theory of the test statistic
in question).
Below we assume $f_0$ is known and remark that our result still
holds provided that $f_0$ can be consistently estimated. In practice,
$f_1$ and $\{\pi_{0i}\}^{m}_{i=1}$ are often unknown and
replaced by their sample counterparts. Let $\hat{f}_1(\cdot)$ and
$\{\hat{\pi}_{0i}\}^{m}_{i=1}$ be the estimators of $f_1(\cdot)$
and $\{\pi_{0i}\}^{m}_{i=1}$ respectively. Define
\begin{equation*}%\label{estimate-lfdr}
\widehat{\mbox{Lfdr}}_i(x) =
\frac{\hat{\pi}_{0i}{f}_0(x)}{\hat{\pi}_{0i}{f}_0(x) +
(1-\hat{\pi}_{0i})\hat{f}_1(x)} =
\frac{\hat{\pi}_{0i}{f}_0(x)}{\hat{f}^i(x)},
\end{equation*}
where $\hat{f}^i(x)=\hat{\pi}_{0i}{f}_0(x) +
(1-\hat{\pi}_{0i})\hat{f}_1(x).$ A natural estimate of $\lambda_m$
can be obtained through
\begin{equation*}%\label{hat-lambda}
\hat{\lambda}_m = \sup \left\{\lambda\in [0,1]:
\frac{\sum_{i=1}^{m}\widehat{\mbox{Lfdr}}_i(x_i)\mathbf{I}
\{\widehat{\mbox{Lfdr}}_i(x_i) \le \lambda
\}}{\sum_{i=1}^{m}\mathbf{I}\{\widehat{\mbox{Lfdr}}_i(x_i) \le
\lambda \}} \le \alpha\right\}.
\end{equation*}
%We show that $\hat{\lambda}_m$ consistently estimates $\lambda_m$ under certain conditions.
Reject the $i$th hypothesis if $\widehat{\text{Lfdr}}_i(x_i)\leq
\hat{\lambda}_m$. This is equivalent to the following step-up
procedure that was originally proposed in \cite{sc07}. Let $\widehat{\mbox{Lfdr}}_{(1)}\leq \cdots \leq
\widehat{\mbox{Lfdr}}_{(m)}$ be the order statistics of $\{
\widehat{\mbox{Lfdr}}_1(x_1), \ldots,
\widehat{\mbox{Lfdr}}_m(x_m)\}$ and denote by $H^{(1)},\dots,
H^{(m)}$ the corresponding ordered hypotheses. Define
\begin{eqnarray*}%\label{step-up}
&&
\hat{k}:=\max\left\{1\leq i\leq m:\frac{1}{i}\sum^{i}_{j=1}\widehat{\mbox{Lfdr}}_{(j)}\leq \alpha\right \};\nonumber \\
&&\mbox{then reject all} \ H^{(i)}\text{ for } i =1,\ldots,\hat{k}.
\end{eqnarray*}
We show that this step-up procedure provides asymptotic control on
the FDR. To facilitate the derivation, we
make the following additional assumption.
\begin{enumerate}
%\item [(C3)] There exist constants $a_1,a_2$ such that
%\begin{align*}
%0<a_1\le \min_{1\leq i\leq m}\pi_{0i}\leq \min_{1\leq i\leq m}\pi_{0i}\le a_2 <1.
%\end{align*}
%Further assume that $f_0$ is uniformly bounded way from zero.

%\item [(C3)] Suppose $D_0(\lambda)$ in %(C1) is differentiable with the %derivative that is uniformly bounded %from above on the interval %$[\lambda_{\infty}/2,1]$.

\item [(C3)] Assume that
$$ \frac{1}{m}\sum^{m}_{i=1}|\widehat{\text{Lfdr}}_i(x_i)-\text{Lfdr}_i(x_i)|\rightarrow^p 0.$$
\end{enumerate}
\begin{remark}

%Condition (C3) ensures that $f^i(x)$ is %bounded way from zero uniformly over both %$x$ and $i,$ which simplifies the %analysis when bounding the difference %between $\mbox{Lfdr}_{i}(x)$ and %$\widehat{\mbox{Lfdr}}_{i}(x)$. Condition

(C4) imposes uniform (weak) consistency on the estimators and Condition (C5) requires the joint empirical distribution function of $\mbox{Lfdr}_{i}(x_i)$ and $x_i$ to converge to a continuous function. Both conditions are useful
in showing that the empirical distribution functions based on $\{\mbox{Lfdr}_{i}(x_i)\}$ and $\{\widehat{\mbox{Lfdr}}_{i}(x_i)\}$ are uniformly close, which is a key step in the proof of Theorem \ref{consistency}. Notice that we do not require the consistency of the estimators at the boundaries. Such a relaxation is important when $\hat{\pi}_{0i}$ and $\hat{f}_1$ are estimated using the shape-restricted approach. For example, \cite{dkl} showed uniform consistency for the Grenander-type estimators on the interval $[\beta_m,1-\beta_m]$ with $\beta_m\rightarrow 0$.
\end{remark}

%By the strong law of large numbers, we have
%$$
%\Big |\frac{1}{m}\sum_{i=1}^{m}\mbox{Lfdr}_i(x_i)I\{\mbox{Lfdr}_i(x_i) \le \lambda\} - \frac{1}{m}\sum_{i=1}^{m}\pi_{0i}F_0(g(a_i \frac{\lambda}{1-\lambda})) \Big | = o_{a.s}(1),
%$$

%We have
%\begin{equation}
%\mbox{aFDR}_m = \frac{\sum_{i=1}^{m} \{ \pi_{0i}F_0(g(\frac{1-\pi_{0i}}{\pi_{0i}}\frac{\lambda}{1-\lambda})) \}}{\sum_{i=1}^{m} \{ \pi_{0i}F_0(g(\frac{1-\pi_{0i}}{\pi_{0i}}\frac{\lambda}{1-\lambda})) + (1-\pi_{0i})F_1(g(\frac{1-\pi_{0i}}{\pi_{0i}}\frac{\lambda}{1-\lambda}))\}}.
%\end{equation}
%At significance level $\alpha,$ we define the oracle critical value $\lambda_{\infty}$ as
%\begin{equation}
%\lambda_{\infty} = \mbox{max}\{\lambda:  \mbox{aFDR}_m\le \lambda\}.
%\end{equation}

%In (C3), we assume $D_0(\lambda)$ has a %bounded derivative on the interval %$[\lambda_{\infty}/2,1]$. This assumption %is not very strong as we still allow its %derivative to be unbounded around zero.

(C3) requires the Lfdr estimators to be consistent in terms of the empirical $L_1$ norm. We shall justify Condition (C3) in Section \ref{sec:c4}.

\begin{theorem}\label{consistency}
Under Conditions (C1)-(C3),
\begin{equation*}
\limsup_{m\rightarrow \infty}\mbox{FDR}_m(\hat{\lambda}_m)\le
\alpha.
\end{equation*}
\end{theorem}
Theorem \ref{consistency} indicates that we can obtain asymptotic
control on the FDR using the data-adaptive procedure when relevant consistent
estimates are available. Similar algorithm has been obtained in
\cite{sc07}, where it is assumed that the hypotheses are exchangeable
in the sense that
$\pi_{01}=\cdots =\pi_{0m}.$ %It is worth mentioning that our proof is based on some new arguments that are different from those in \cite{sc07}.

%The proofs of Theorem \ref{fdr-control} and Theorem \ref{consistency} use Berry-Esseen bound \cite{pena-lai-shao} and are quite different from those in \cite{sc07}, where marginal
%FDR is controlled.
%For statistical inference, we need to obtain estimates of $\pi_{0i}, i =1, \ldots, m$ and ${f}_1(\cdot)$ that satisfy conditions specified in Theorem \ref{consistency}.

%\begin{rem}
%{\rm Under the assumption that $f_1$ and $\hat{f}_1$ are both decreasing (see Section \ref{general:alg}), $f_1$ is continuous, and
%$$|\hat{f}_1(x)-f_1(x)| \overset{p}{\rightarrow} 0,$$
%for any $x>0$, it can be shown that
%\begin{align}\label{unif}
%\sup_{x\in [\epsilon,1]}|\hat{f}_1(x)-f_1(x)|\overset{p}{\rightarrow} 0
%\end{align}
%for any $\epsilon>0$ using similar arguments as in the proof of
%Glivenko-Cantelli Theorem \cite{vw1996}. %If we additionally assume that
%$f_1(0)<\infty$, then $$\sup_{x\in [0,1]}|\hat{f}_1(x) - f_1(x)
%|\overset{p}{\rightarrow} 0,$$
%which justifies the second part of (C4).
%}
%\end{rem}

\section{Estimating the unknowns}\label{sec:iso}
%For statistical inference, we need to obtain estimates of
%$\{\pi_{0i}\}^{m}_{i=1}$ and ${f}_1(\cdot)$ that satisfy conditions
%specified in Theorem \ref{consistency}.

\subsection{The density function $f_1(\cdot)$ is known}\label{subsec:consis}
We first consider the case that $f_0(\cdot)$ and $f_1(\cdot)$ are both known. Under such setup, we need to estimate $m$ unknown parameters $\pi_{0i}, i =1, \ldots, m,$ which is prohibitive without additional constraints. One constraint that makes the problem solvable is the monotone constraint. In statistical
genetics and genomics, investigators can use auxiliary information (e.g., $p$-values
from previous or related studies) to generate a ranked list of hypotheses $H_1,\dots,H_m$
even before performing the experiment, where $H_1$ is the hypothesis that the investigator believes
to most likely correspond to a true signal, while $H_m$ is the one believed to be least likely.
Specifically, let $\Pi_0 = (\pi_{01}, \ldots, \pi_{0m})\in (0,1)^m.$ Define the convex set
$$
{\cal M} = \{ \Pi = (\pi_{1}, \ldots, \pi_m) \in (0, 1)^m: 0\leq \pi_1 \le \ldots \le \pi_m \leq 1\}.
$$
We illustrate the motivation for the monotone constraint with an example.
\begin{example}
{\rm
Suppose that we are given data consisting of a pair of values $(x_{i1},x_{i2}),$ where $x_{i1}$ represents the $p$-value, $x_{i2}$ represents auxiliary information
and they are independent conditional on the hidden true state $\theta_i$ for
$i=1,\ldots,m.$ Suppose
\begin{equation}\label{data-generation}
x_{ij} \mid \theta_i \overset{\text{ind}}{\sim} (1-\theta_i)f_{0,j}(x_{ij}) + \theta_i f_{1,j}(x_{ij}), \quad \quad i = 1, \ldots, m, \quad j =1, 2,
\end{equation}
where $\theta_i =1$ if $H_i$ is alternative and $\theta_i=0$ if $H_i$ is null, $f_{0,j}(\cdot)$ is the density function of $p$-values or auxiliary variables under the null hypothesis and $f_{1,j}(\cdot)$ is the density function of $p$-values or auxiliary variables under the alternative hypothesis. Suppose $P(\theta_i=0)=\tau_0$ for all $i =1, \ldots, m.$ Using the Bayes rule and the independence between $x_{i1}$ and $x_{i2}$ given $\theta_i, i =1, \ldots, m,$ we have the conditional distribution of $x_{i1}\mid x_{i2}$ as follows:
\begin{align*}
&f(x_{i1} \mid x_{i2})
\\=& \frac{f(x_{i1}, x_{i2} \mid \theta_i = 0)\tau_0 + f(x_{i1}, x_{i2} \mid \theta_i = 1)(1-\tau_0)}{f(x_{i2} \mid \theta_i = 0 )\tau_0 + f(x_{i2}\mid \theta_i = 1)(1-\tau_0)}\\
 =& \frac{f(x_{i1} \mid \theta_i = 0)f(x_{i2}\mid \theta_i = 0)\tau_0 + f(x_{i1}\mid \theta_i = 1)f(x_{i2}\mid \theta_i=1)(1-\tau_0)}{f(x_{i2}\mid \theta_i = 0)\tau_0 + f(x_{i2}\mid \theta_i = 1)(1-\tau_0)}\\
=& \frac{f_{0,1}(x_{i1})f_{0,2}(x_{i2})\tau_0 + f_{1,1}(x_{i1})f_{1,2}(x_{i2})(1-\tau_0)}{f_{0,2}(x_{i2})\tau_0 + f_{1,2}(x_{i2})(1-\tau_0)}\\
=& f_{0,1}(x_{i1})\gamma_0(x_{i2}) + f_{1,1}(x_{i1})(1-\gamma_0(x_{i2})),
\end{align*}
where
$$\gamma_0(x) = \frac{f_{0,2}(x)\tau_0}{f_{0,2}(x)\tau_0 + f_{1,2}(x)(1-\tau_0)} = \frac{\tau_0}{\tau_0 + \frac{f_{1,2}(x)}{f_{0,2}(x)}(1-\tau_0)}.$$
If $f_{1,2}(x)/f_{0,2}(x)$ is a monotonic function, so is $\gamma_0(x)$.
Therefore, the order of $x_{i2}$ generates a ranked list of the hypotheses $H_1,\dots,H_m$ through the conditional prior probability $\gamma_0(x)$.
}
\end{example}

We estimate $\Pi_0$ by solving the following maximum likelihood problem:
\begin{equation}\label{opt-orl}
\begin{split}
&\hat{\Pi}_0=(\hat{\pi}_{01},\dots,\hat{\pi}_{0m})=\argmax_{\Pi=(\pi_{1},\dots,\pi_{m})\in\mathcal{M}}l_m(\Pi),\\
&l_m(\Pi):=\sum_{i=1}^{m}\log\left\{\pi_{i}f_0(x_i)+(1-\pi_{i})f_1(x_i)\right\}.
\end{split}
\end{equation}
It is easy to see that (\ref{opt-orl}) is a convex optimization
problem. Let
$\phi(x,a)=af_0(x)+(1-a)f_1(x)$. To facilitate the derivations, we
shall assume that $f_0(x_i)\neq f_1(x_i)$ for all $i$, which is a
relatively mild requirement. Under this assumption, it is straightforward to see
that for any $1\leq k\leq l\leq m$, $\sum^{l}_{i=k}\log\phi(x_i,a)$
is a strictly concave function for $0< a < 1.$ Let
$\hat{a}_{kl}=\argmax_{a\in [0,1]} \sum^{l}_{i=k}\log \phi(x_i,a)$ be the
unique maximizer. According to Theorem 3.1 of \cite{rw}, we have
\begin{align}\label{eq-maxmin}
\hat{\pi}_{0i}=\max_{1\leq k\leq i}\min_{i\leq l\leq m}\hat{a}_{kl}.
\end{align}
However, this formula is not practically useful due to the computational burden when $m$ is very large.
Below we suggest a more efficient way to solve problem (\ref{opt-orl}). A
general algorithm when $f_1$ is unknown is provided in the next subsection. %Section
%\ref{general:alg}. 
%\textcolor{green}
{The main computational tools are the EM algorithm for two-group mixture model and the Pool-Adjacent-Violator-Algorithm from isotonic regression for the monotone constraint on the prior probability of null hypothesis $\pi_{0i}, i =1, \ldots, m$ \cite{D77, rwd1988}. Our procedure is tuning parameter free and can be easily implemented in practice. The EM algorithm treats the hidden state $\theta_i, i =1, \ldots, m$ as missing data. The isotonic regression problem is to 
\begin{equation}\label{isto}
\begin{split}
&\mbox{minimize} \quad \sum_{i=1}^{m}(a_i - z_i )^2w_i,\\
&\mbox{subject to} \quad z_1 \leq z_2\leq \ldots \leq z_m, 
\end{split}
\end{equation}
where $w_i>0$ and $a_{i}, i =1, \ldots, m$ are given. % and $\preceq$ is a specified ordering on $\{1, \ldots, m\}.$ 
By   \cite{benzi2020}, the solution to (\ref{isto}) can be written as 
 \begin{equation}\label{hatz}
 \hat{z}_i = \max_{a \leq i}\min_{b \geq i}\frac{\sum_{j=a}^{b}a_jw_j}{\sum_{j=a}^{b}w_j}.
 \end{equation}
We need a key result from \cite{bb72} which we present below for completeness.}
%\textcolor{green}
{\begin{proposition}(Theorem 3.1 in \cite{bb72}) \label{prop-dual}
Let $G$ be a proper convex function on $R,$ and $g$ its derivative. Denote $m$ dimensional vectors $\mathbf{a} = (a_1, \ldots, a_m)$ and $\mathbf{w} = (w_1, \ldots, w_m).$
%Let $a$ and $w>0$ be given functions on $\Omega = \{1, 2, \ldots, m \},$ regarded as points of 
$R^m: \mathbf{a} = (a_1, \ldots, a_m), \mathbf{w} = (w_1, \ldots, w_m).$ %Let $G$ be a proper convex function on $R,$ $g$ its derivative and $K$ a convex cone of isotonic functions on $\Omega$ having range in $R.$ 
We call the problem
\begin{equation}\label{g-istonic}
\mbox{minimize}_{z=(z_1, \ldots, z_m)}\sum_{i=1}^{m}\{G(z_i) - a_iz_i \}w_i
\end{equation}
the generalized isotonic regression problem. %Suppose that the range of $\mathbf{a}$ is in the effective domain of $g^{-1}.$ %where $\mathbf{a}$ is a given function on $\Omega.$ 
Then 
\begin{equation}
z_i^{\circ} = g^{-1}(a_i^*), \quad i = 1, \ldots, m,
\end{equation} 
where 
\begin{equation}\label{hatastar}
a_i^* = \max_{a\leq i}\min_{b \geq i}\frac{\sum_{j=a}^{b}a_jw_j} {\sum_{j=a}^{b}w_j},
\end{equation}
solves the generalized isotonic regression problem (\ref{g-istonic}). %and 
%\begin{equation}\label{hatastar}
%a_i^* = \max_{a\leq i}\min_{b \geq i}\frac{\sum_{j=a}^{b}a_jw_j} {\sum_{j=a}^{b}w_j}.
%\end{equation}
The minimization function is unique if $G$ is strictly convex.
\end{proposition}}

%\textcolor{green}
{Observe that (\ref{hatz}) and (\ref{hatastar}) have the same expression. In practice, we implement the Pool-Adjacent-Violator-Algorithm by solving (\ref{isto}) to get $(\ref{hatastar}).$}
%We will use this key result twice, first in estimating $\pi_{0i}, i =1, \ldots, m$ when $f_1$ is known and second in estimating both $\pi_{0i}, i=1, \ldots, m$ and $f_1$ with double monotone constraint.}

%\textcolor{green}
{ Let
$\Pi^{(t)}=(\hat{\pi}_{01}^{(t)},\dots,\hat{\pi}_{0m}^{(t)})$ be the
solution at the $t$th iteration. Define
\begin{align*}
&Q_{j}^{(t)}:=Q_j^{(t)}(\hat{\pi}_{0j}^{(t)})=\frac{\hat{\pi}_{0j}^{(t)}f_0(x_j)}{\hat{\pi}_{0j}^{(t)}f_0(x_j)+(1-\hat{\pi}_{0j}^{(t)})f_1(x_j)},\\
&Q(\Pi|\Pi^{(t)})=\sum^{m}_{j=1}\{Q^{(t)}_{j}\log(\pi_j)+(1-Q^{(t)}_j)\log(1-\pi_j)\}.
\end{align*}
At the $(t+1)$th iteration, we solve the following problem,
\begin{align}\label{iso-0}
\Pi^{(t+1)}=\argmax_{\Pi=(\pi_{1},\dots,\pi_{m})\in\mathcal{M}}Q(\Pi|\Pi^{(t)}).
\end{align}
To use Proposition \ref{prop-dual}, we first do a change of variable by letting $z_i = \log \pi_{0i}, i =1, \ldots, m.$ We proceed by solving
\begin{eqnarray*}
\argmin_{z_1\le z_2\le\cdots \le  z_m}\sum_{i=1}^{m}(1-Q_i^{(t)})\{-\log (1-e^{z_i}) - \frac{Q_{i}^{(t)}}{1-Q_i^{(t)}} z_i\}.
%\\=&\arg\min_{\Pi\in\mathcal{M}}\sum^{m}_{i=1}\left(Q_i^{(t)}-\pi_i\right)^2,\\
%f_1^{(t+1)}=&\arg\max_{\tilde{f}_1\in\mathcal{H}}\sum^{m}_{i=1}(1-Q_i^{(t)})\log \tilde{f}_1(x_i).
\end{eqnarray*}
%where $z_i = \log \pi_{0i}, i =1, \ldots, m.$
%By Proposition \ref{prop-dual}, we have
%$$
%\hat{z}_i = \max_{a \leq i} \min_{b \geq i}\frac{\sum_{j=a}}{b}Q_j^{(t)}{\sum_{j=a}^{b}(1-Q_i^{(t)})},
%$$
which has the same solution to the problem
\begin{eqnarray*}
\argmin_{z_1 \le z_2 \le \ldots \le z_m}\sum_{j=1}^{m}\{\frac{Q_i^{(t)}}{1-Q_i^{(t)}} -z_i\}^2(1-Q_i^{(t)}).
\end{eqnarray*}
We write it as 
\begin{eqnarray*}
\hat{z}_i &=& \max_{a \leq i} \min_{b \geq i}\frac{\sum_{j=a}^{b}Q_j^{(t)}}{\sum_{j=a}^{b}(1-Q_j^{(t)})},  \\ %\quad i =1, \ldots, m.
&=& \frac{\max_{a\leq i}\min_{b\geq i} \frac{1}{b-a+1}\sum_{j=a}^{b}Q_j^{(t)}}{1-\max_{a\leq i}\min_{b \geq i}\frac{1}{b-a+1}\sum_{j=a}^{b}Q_j^{(t)}}, \quad i = 1, \ldots, m.
\end{eqnarray*}
By Proposition \ref{prop-dual}, we have 
$$
\hat{\pi}_{0i}^{(t+1)} = \frac{\hat{z}_i}{\hat{z}_i + 1} = \max_{a \leq i}\min_{b \geq i} \frac{1}{b-a+1}\sum_{j=a}^{b}Q_j^{(t)}, \quad i =1, \ldots, m. 
$$
%$$
%\hat{\pi}_{0i}^{(t+1)} = \frac{\max_{a \leq i} \min_{b \geq i}\frac{\sum_{j=a}^{b}Q_i^{(t)}}{\sum_{j=a}^{b}(1-Q_i^{(t)})}}{\max_{a\leq i} \min_{b \geq i}\frac{\sum_{j=a}^{b}Q_i^{(t)}}{\sum_{j=a}^{b}(1-Q_i^{(t)})} +1}
%$$
}
%where $
%{\cal M} = \{ \Pi = (\pi_{1}, \ldots, \pi_m) \in (0, 1)^m: 0\leq \pi_1 \le \ldots \le \pi_m \leq 1\}.$}

%Let
%$\Pi^{(t)}=(\hat{\pi}_{01}^{(t)},\dots,\hat{\pi}_{0m}^{(t)})$ be the
%solution at the $t$th iteration. Define
%\begin{align*}
%&Q_{j}^{(t)}:=Q_j^{(t)}(\hat{\pi}_{0j}^{(t)})=\frac{\hat{\pi}_{0j}^{(t)}f_0(x_j)}{\hat{\pi}_{0j}^{(t)}f_0(x_j)+(1-\hat{\pi}_{0j}^{(t)})f_1(x_j)},\\
%&Q(\Pi|\Pi^{(t)})=\sum^{m}_{j=1}\{Q^{(t)}_{j}\log(\pi_j)+(1-Q^{(t)}_j)\log(1-\pi_j)\}.
%\end{align*}
%At the $(t+1)$th iteration, we solve the following problem,
%\begin{align}\label{iso-0}
%\Pi^{(t+1)}=\argmax_{\Pi=(\pi_{1},\dots,\pi_{m})\in\mathcal{M}}Q(\Pi|\Pi^{(t)}).
%\end{align}
%By Theorem 3.1 of \cite{bb72}, we only need to solve the isotonic regression
%problem,
%\begin{align}\label{iso-1}
%&\hat{\Pi}^{(t+1)}=\argmin_{\Pi=(\pi_{1},\dots,\pi_{m})\in\mathcal{M}}
%\sum_{j=1}^{m}\left\{Q_{j}^{(t)}-\pi_{j}\right\}^2.
%\end{align}
%The solution to (\ref{iso-1}) has an explicit form given by the
%max-min formula,
%\begin{align*}
%\hat{\pi}_{0i}^{(t+1)}=\max_{a\leq i}\min_{b\geq i}\frac{\sum^{b}_{j=a}Q_{j}^{(t)}}{b-a+1},
%\end{align*}
%\textcolor{green}
{In practice, we obtain $\hat{\pi}_{0i}^{(t+1)}, i =1, \ldots, m$ through the Pool-Adjacent-Violators Algorithm (PAVA) \cite{rwd1988}} by solving the following problem
\begin{equation} \label{iso-2}
\Pi^{(t+1)} = \argmin_{\Pi = (\pi_1, \ldots, \pi_m)\in \mathcal {M}}\sum_{j=1}^{m}\{Q_j^{(t)} -\pi_j \}^2.
\end{equation}
%which can be obtained conveniently using the Pool-Adjacent-Violators
%Algorithm (PAVA) \cite{rwd1988}. 
Note that if $Q^{(t)}_1\geq Q^{(t)}_2\geq \cdots Q^{(t)}_m$, then the solution to (\ref{iso-2}) is simply given by $\hat{\pi}_{0i}^{(t+1)}=\sum^{m}_{j=1}Q^{(t)}_j/m$ for all $1\leq i\leq m.$ 
As the EM algorithm is a hill-climbing algorithm, it is not hard to show that $l_m(\Pi^{(t)})$
is a non-decreasing function of $t$. %See Lemma \ref{lemma1} in the appendix.

We study the asymptotic consistency of the true maximum likelihood estimator $\hat{\Pi}_0$ which can be represented as (\ref{eq-maxmin}). To this end, consider the model
$$x_i \overset{\text{i.i.d}}{\sim} \pi_{0i}f_0+(1-\pi_{0i})f_1,\quad \pi_{0i}=\pi_0(i/m)\footnote{For the ease of presentation,
we suppress the dependence on $m$ in $\pi_{0i}$.},$$ for some
non-decreasing function $\pi_0:[0,1]\rightarrow [0,1].$ Our first
result concerns the point-wise consistency for each $\hat{\pi}_{0i}$.
For a set $A$, denote by $\text{card}(A)$ its cardinality.
\begin{theorem}\label{thm1}
Assume that $\int (\log f_i(x))^2f_j(x)dx<\infty$
for $i,j=0,1$, and
$P(f_0(x_i)=f_1(x_i))=0$. Suppose $0<\pi_0(0)\leq \pi_0(1)<1.$ For
any $\epsilon>0$, let $0\leq t'<i_0/m<t''\leq 1$ such that
$|\pi_0(t')-\pi_0(i_0/m)|\vee |\pi_0(t'')-\pi_0(i_0/m)|<\epsilon/2.$
Denote $A_1=\{i:t'\leq i/m\leq i_0/m\}$ and $A_2=\{i:i_0/m\leq
i/m\leq t''\}$. For $\text{card}(A_1)\wedge \text{card}(A_2)\geq N,$ we
have
$$P\left(
|\hat{\pi}_{0,i_0}-\pi_{0,i_0}|<\epsilon\right)\geq
1-O\left(\frac{1}{\epsilon^2N}\right).$$
\end{theorem}
The condition on the cardinalities of $A_1$ and $A_2$ guarantees
that there are sufficient observations around $i_0/m$, which allows
us to borrow information to estimate $\pi_{0,i_0}$ consistently. The assumption
$P(f_0(x_i)=f_1(x_i))=0$ ensures that the maximizer
$\hat{a}_{kl}$ is unique for $1\leq k\leq l \leq m.$ It is fulfilled if the set $\{x\in [0,1]: f_0(x)=f_1(x)\}$ has zero Lebesgue measure. As
a direct consequence of Theorem \ref{thm1}, we have the following
uniform consistency result of $\hat{\Pi}_0$. Due to the monotonicity, the uniform convergence follows from the pointwise convergence.
\begin{corollary}\label{cor1}
For $\epsilon>0,$ suppose there exists a set $i_1<i_2<\cdots<i_l$,
where each $i_k$ satisfies the assumption for $i_0$ in Theorem
\ref{thm1} and that $\max_{2\leq k\leq
l}(\pi_{0,i_k}-\pi_{0,i_{k-1}})<\epsilon.$ Then we have
\begin{align*}
P\left( \max_{i_1\leq i\leq
i_l}|\hat{\pi}_{0,i}-\pi_{0,i}|<\epsilon\right)\geq
1-O\left(\frac{l}{\epsilon^2N}\right).
\end{align*}
\end{corollary}
\begin{rem}
{\rm
Suppose $\pi_0$ is Lipschitz continuous with the Lipschitz constant $K$. Then we can set $t''=(i_0-1)/m+\epsilon/(2K)$, $t'=(i_0+1)/m-\epsilon/(2K)$ and thus $N=\lfloor m\epsilon/(2K)\rfloor$. Our result suggests that
$$P\left(
|\hat{\pi}_{0,i_0}-\pi_{0,i_0}|<\epsilon\right)\geq
1-O\left(\frac{K}{\epsilon^3 m}\right),$$
which implies that $|\hat{\pi}_{0,i_0}-\pi_{0,i_0}|=O_p(m^{-1/3}).$
}
\end{rem}

\subsection{The density function $f_1(\cdot)$ is unknown}\label{general:alg}
In practice, $f_1$ and $\Pi_0$ are both unknown. We propose to estimate
$f_1$ and $\Pi_0$ by maximizing the likelihood, i.e.,
\begin{equation}\label{m4}
(\hat{\Pi}_0,\hat{f}_1)=\argmax_{\Pi\in\mathcal{M},\tilde{f}_1\in\mathcal{H}}\sum_{i=1}^{m}\log\left\{\pi_{i}f_0(x_i)+(1-\pi_{i})\tilde{f}_1(x_i)\right\},
\end{equation}
where $\mathcal{H}$ is a pre-specified class of density functions. %\textbf{Jun: can you provide some details about the choice of $\varrho_1$ and $\varrho_2$ in your algorithm?}
In (\ref{m4}),
$\mathcal{H}$ might be the class of beta mixtures or the class of
decreasing density functions. Problem (\ref{m4}) can be solved by
Algorithm 1. A derivation of Algorithm 1 from the full data likelihood
that has access to latent variables is provided in the Appendix.
Our algorithm is quite general in the sense that it
allows users to specify their own updating scheme for the density
components in (\ref{em-2}). Both parametric and non-parametric
methods can be used to estimate $f_1$.

\begin{algorithm}[H]\label{alg11}\caption{}
\begin{flushleft}
0. Input the initial values $(\Pi^{(0)},f_1^{(0)})$.\\
1. \textbf{E-step:} Given $(\hat{\Pi}^{(t)},\hat{f}_1^{(t)})$, let
  $$Q_{i}^{(t)}=\frac{\hat{\pi}_{0i}^{(t)}f_0(x_i)}{\hat{\pi}_{0i}^{(t)}f_0(x_i)+(1-\hat{\pi}_{0i}^{(t)})\hat{f}_1^{(t)}(x_i)}.$$
2. \textbf{M-step:} Given $Q_{i}^{(t)}$, update $(\Pi,f_1)$ through
\begin{equation}\label{em-1}
\begin{split}
&(\hat{\pi}_{01}^{(t+1)},\dots,\hat{\pi}_{0m}^{(t+1)})=\argmin_{\Pi=(\pi_{1},\dots,\pi_{m})\in\mathcal{M}}
\sum_{i=1}^{m}\left(Q_{i}^{(t)}-\pi_{i}\right)^2,
\end{split}
\end{equation}
and
\begin{align}\label{em-2}
\hat{f}_1^{(t+1)}=\argmax_{\tilde{f}_1\in
\mathcal{H}}\sum^{m}_{i=1}(1-Q_{i}^{(t)})\log \tilde{f}_1(x_i).
\end{align}
3. Repeat the above E-step and M-step until the algorithm converges.
\end{flushleft}
\end{algorithm}

In the multiple testing literature, it is common to assume that
$f_1$ is a decreasing density function (e.g., smaller $p$-values imply stronger
evidence against the null), see e.g. \cite{langaas2005}. As an
example of the general algorithm, let $\mathcal{H}$ denote the
class of decreasing density functions. We shall discuss how (\ref{em-2}) can
be solved using the PAVA. %\textcolor{green}
{The key recipe is to use Proposition \ref{prop-dual} in obtaining $f_1$ evaluated at the observed $p$-values.}
%\textbf{The first question to address is the identifiability of the model:
%$$f(z|t,\pi_0,f_1):=\pi_0(t)f_0(z)+(1-\pi_0(t))f_1(z),$$
%where $f_1(z)$ is a decreasing density (which is independent of $t$) and $0\leq \pi_0(t)\leq 1$ is increasing in $t$.
%If $f(z|t,\pi_0,f_1)=f(z|t,\pi_0^*,f_1^*)$ implies that $\pi_0=\pi_0^*$ and $f_1=f_1^*$, then the model is identifiable.}
%When the model is identifiable,  to estimate $\pi_0$ and $f_1$ in the following way. Specifically, we update $f_1$ in the EM algorithm through:
%\begin{align}\label{obj-shape}
%f_1^{(t+1)}=\argmax_{f_1} \sum^{m}_{i=1}(1-Q_{i}^{(t)})\log f_1(z_i),
%\end{align}
%subject to the shape constraint that $f_1$ is a decreasing density function on $[0,1]$.
Specifically, it can be accomplished by a series of steps outlined
below. Define the order statistics of $\{x_i\}$ as $x_{(1)}\leq x_{(2)}\leq \cdots \leq x_{(m)}.$ Let $Q_{(i)}^{(t)}$ be the corresponding $Q_i^{(t)}$ that is associated with $x_{(i)}$.
\\ \emph{Step 1:} The objective function in (\ref{em-2}) only looks at the value of $f_1$ at $x_{(i)}$. The objective function increases if $f_1(x_{(i)})$ increases, and
the value of $f_1$ at $(x_{(i-1)},x_{(i)})$ has no impact on the objective
function (where $x_{(0)}=0$). Therefore, if $f$ maximizes the objective
function, there is a solution that is constant on $(x_{(i-1)},x_{(i)}]$.
\\ \emph{Step 2:} Let $y_i=f_1(x_{(i)})$. We only need to find $y_i$ which maximizes
$$\sum^{m}_{i=1}(1-Q_{(i)}^{(t)})\log (y_i),$$
subject to $y_1\geq y_2\geq \cdots \geq y_m\geq 0$ and
$\sum_{i=1}^{m}y_i(x_{(i)}-x_{(i-1)})=1$. It can be formulated as a convex
programming problem which is tractable. In Steps 3 and 4 below, we further translate it into an isotonic regression problem.
%\textbf{Question: Can we apply the PAVA in this case? Below are some thoughts. Please let me know if it makes sense.}
\\ \emph{Step 3:} Write $Q^{(t)}=\sum^{m}_{i=1}(1-Q_{(i)}^{(t)})$. Consider the problem:
$$\min\sum^{m}_{i=1}\left\{-(1-Q_{(i)}^{(t)})\log (y_i)+Q^{(t)}y_i(x_{(i)}-x_{(i-1)})\right\}.$$
The solution is given by
$\hat{y}_i=\frac{1-Q_{(i)}^{(t)}}{Q^{(t)}(x_{(i)}-x_{(i-1)})}$, which
satisfies the constraint $\sum_{i=1}^{m}y_i(x_{(i)}-x_{(i-1)})=1$ in Step
2.
\\ \emph{Step 4:} Rewrite the problem in Step 3 as
$$\min\sum^{m}_{i=1}(1-Q_{(i)}^{(t)})\left\{-\log (y_i)-\frac{-Q^{(t)}(x_{(i)}-x_{(i-1)})}{(1-Q_{(i)}^{(t)})}y_i\right\}.$$ 
%\textcolor{green}
{This is the generalized isotonic regression problem considered in Proposition \ref{prop-dual}. We use (\ref{hatz}) to obtain (\ref{hatastar})} as follows. 
Let
$$(\hat{u}_1,\dots,\hat{u}_m)=\argmin\sum_{i=1}^{m}(1-Q_{(i)}^{(t)})\left(-\frac{Q^{(t)}(x_{(i)}-x_{(i-1)})}{(1-Q_{(i)}^{(t)})}-u_i\right)^2$$
subject to $u_1\geq u_2\geq \cdots\geq u_m.$ The solution is given
by the max-min formula
\begin{align*}
\hat{u}_i = \max_{b\geq i}\min_{a\leq
i}\frac{-Q^{(t)}\sum^{b}_{j=a}(x_{(j)}-x_{(j-1)})}{\sum^{b}_{j=a}(1-Q_{(j)}^{(t)})},
\end{align*}
which can be obtained using the PAVA. %\textcolor{green}
{By Proposition \ref{prop-dual}, we arrive at the solution to the original problem (\ref{em-2}) by letting $\tilde{y}_i=-\frac{1}{\hat{u}_i}.$} %which is the solution to the
%original problem (\ref{em-2}) according to Theorem 3.1 of
%\cite{bb72}. 
Therefore, in the EM-algorithm, one can employ the PAVA
to estimate both the prior probabilities of being null and the $p$-value density function under the alternative hypothesis. Because of this, our algorithm is fast and tuning parameter
free, and is very easy to implement in practice.

\subsection{Asymptotic convergence and verification of Condition (C3)}\label{sec:c4}
In this subsection, we present some convergence results regarding the proposed estimators in Section \ref{general:alg}. Furthermore,
we propose a refined estimator for $\pi_0$, and justify Condition (C3) for the corresponding Lfdr estimator. Throughout the following discussions, we assume that
$$x_i \sim f^i=\pi_0(i/m)f_0+(1-\pi_0(i/m))f_1$$
independently for $1\leq i\leq m$ and $\pi_0:[0,1]\rightarrow [0,1]$ with $\pi_0(i/m)=\pi_{0i}$. Let $\mathcal{F}$ be the class of densities defined on $[0,1]$.
For $f,g\in\mathcal{F}$, we define the squared Hellinger-distance as
\begin{align*}
H^2(f,g)=\frac{1}{2}\int_{0}^{1}(\sqrt{f(x)}-\sqrt{g(x)})^2 dx=1-\int_{0}^{1}\sqrt{f(x)g(x)}dx.
\end{align*}
%Let $\mathcal{S}_m$ be the set
%of all configurations $S_m$ of $m$ (possibly %non-distinct) points within $[0,1]$. For $f$ defined on %$[0,1]$, let $\|f\|_{S_m}=m^{-1}\sum_{x\in S_m} |f(x)|$.
Suppose the true alternative density $f_1$ belongs to a class of decreasing density functions $\mathcal{H}\subset \mathcal{F}$. Let $\Xi=\{\pi:[0,1]\rightarrow [0,1], 0<\varepsilon<\pi(0)\leq \pi(1)<1-\varepsilon<1, \text{ and $\pi(\cdot)$ is nondecreasing}\}$
and assume that $\pi_0\in \Xi.$ Consider $\tilde{f}^i=\tilde{\pi}(i/m)f_0 + (1-\tilde{\pi}(i/m))\tilde{f}_1$ and $\breve{f}^i=\breve{\pi}(i/m)f_0 + (1-\breve{\pi}(i/m))\breve{f}_1$ for $1\leq i\leq m$, $\tilde{f}_1,\breve{f}_1\in \mathcal{H}$ and $\tilde{\pi}, \breve{\pi} \in \Xi$. Define the average squared Hellinger-distance
between $(\tilde{\pi},\tilde{f}_1)$ and $(\breve{\pi},\breve{f}_1)$ as
\begin{align*}
&H^2_m((\tilde{\pi},\tilde{f}_1),(\breve{\pi},\breve{f}_1))=\frac{1}{m}\sum_{i=1}^{m}H^2(\tilde{f}^i,\breve{f}^i).
\end{align*}
Suppose $(\hat{\pi}_0,\hat{f}_1)$ is an estimator of $(\pi_0,f_1)$ such that
$$\sum^{m}_{i=1}\log\left(\frac{2\hat{f}^i(x_i)}{\hat{f}^i(x_i)+f^i(x_i)}\right)\geq 0,$$
where $\hat{f}^i(x)=\hat{\pi}_0(i/m)f_0(x)+(1-\hat{\pi}_0(i/m))\hat{f}_1(x)$.
Note that we do not require $(\hat{\pi}_0,\hat{f}_1)$ to be the global maximizer of the likelihood.
We have the following result concerning the convergence of $(\hat{\pi}_0,\hat{f}_1)$ to $(\pi_0,f_1)$ in terms of the average squared Hellinger-distance.

\begin{theorem}\label{thm:H-dist}
Suppose $\pi_0 \in \Xi$, $f_0\equiv1$, and $f_1\in\mathcal{H}$. Under the assumption that $\int^{1}_{0}f_1^{1+a}(x)dx <\infty$ for some $0<a\leq 1,$
we have
\begin{align*}
P\left(H_m((\pi_0,f_1),(\hat{\pi}_0,\hat{f}_1))> M m^{-1/3}\right)
\leq M_1\exp(-M_2 m^{1/3}),
\end{align*}
for some $M,M_1$ and $M_2>0.$ We remark that $f_1(x)=(1-\gamma)x^{-\gamma}$ with $0<\gamma<1$ satisfies $\int^{1}_{0}f_1^{1+a}(x)dx <\infty$ for $0<a<(1/\gamma-1)\wedge 1$.
\end{theorem}
%\begin{rem}
%{\rm Suppose
%$$\limsup_{x\downarrow 0}\frac{\max_{f\in %\mathcal{H}}f(x)}{1/x^k}<\infty.$$
%for $0<k<1$. Then we can pick $\xi_m=1/(m\log %m)^{1/(1-k)}$
%and thus $\gamma_m=(m \log m)^{k/(1-k)}$. When
%$\tilde{a}>k$, we have $\gamma_m(\gamma_m %m)^{-\tilde{a}}=o(1)$,
%and hence $$H^2_m((\pi_0,f_1),(\hat{\pi}_0,\hat{f}_1))=O%_p((\gamma_m m)^{1-\tilde{a}})=o_p(1).$$
%}
%\end{rem}
Theorem \ref{thm:H-dist} follows from an application of Theorem 8.14 in \cite{van2000}. %\textcolor{green}
{By Cauchy-Schwarz inequality,} it is known that
\begin{align*}
\int^{1}_{0}|f(x)-g(x)|dx %=&\sqrt{\int^{1}_{0}(\sqrt{f(x)}-\sqrt{g(x)})^2dx}\sqrt{\int^{1}_{0}(\sqrt{f(x%)}+\sqrt{g(x)})^2dx}
\leq& 2H(f,g)\sqrt{2-H^2(f,g)}.
\end{align*}
Under the conditions in Theorem \ref{thm:H-dist}, we have
\begin{align}\label{eq-f-con}
\frac{1}{m}\sum^{m}_{i=1}\int^{1}_{0}|\hat{f}^i(x)-f^i(x)|dx=O_p(m^{-1/3}).
\end{align}
However, $\pi_0$ and $f_1$ are generally unidentifiable without extra conditions.
Below we focus on the case $f_0\equiv1$. The model is identifiable in this case if there exists an $a_0\leq 1$ such that $f_1(a_0)=0$. If $f_1$ is decreasing, then $f_1(x)=0$ for $x\in [a_0,1]$. Suppose $a_0<1$. For a sequence $b_m\in (0,1)$ such that
\begin{align}\label{rate}
\frac{\int^{1}_{b_m}f_1(x)dx}{1-b_m}=o(1),\quad \frac{m^{-1/3}}{1-b_m}=o(1),
\end{align}
as $m\rightarrow+\infty$, we define the refined estimator for $\pi_0(i/m)$ as
\begin{align*}
\breve{\pi}_0(i/m)=\frac{1}{1-b_m}\int^{1}_{b_m} \hat{f}^i(x)dx=\hat{\pi}_{0}(i/m)+(1-\hat{\pi}_{0}(i/m))\frac{\int^{1}_{b_m}\hat{f}_1(x)dx}{1-b_m}.
\end{align*}
Under (\ref{rate}), we have
\begin{equation}\label{eq-pi-con}
\begin{split}
&\frac{1}{m}\sum^{m}_{i=1}\left|\breve{\pi}_0(i/m)-\pi_0(i/m)\right|
\\= & \frac{1}{m(1-b_m)}\sum^{m}_{i=1}\left|\int^{1}_{b_m}\hat{f}^i(x)dx-\int^{1}_{b_m}f^i(x)dx\right|+o_p(1)
\\ \leq&
\frac{1}{m(1-b_m)}\sum^{m}_{i=1}\int^{1}_{0}|\hat{f}^i(x)-f^i(x)|dx+o_p(1)=o_p(1).
\end{split}
\end{equation}
Given the refined estimator $\breve{\pi}_0$, the Lfdr can be estimated by
\begin{align*}
\widehat{\text{Lfdr}}_i(x_i)=\frac{\breve{\pi}_0(i/m)}{\hat{f}^i(x_i)}.
\end{align*}
As $\hat{\pi}_0,\pi_0\in \Xi$ and thus are bounded from below, by (\ref{eq-f-con}) and (\ref{eq-pi-con}), it is not hard to show that
\begin{align}\label{eq-fdr-con}
\frac{1}{m}\sum^{m}_{i=1}\int^{1}_{0}|\widehat{\text{Lfdr}}_i(x)-\text{Lfdr}_i(x)|dx=o_p(1).
\end{align}
Moreover, we have the following result which justifies Condition (C3).
\begin{corollary}\label{cor-c4}
{\rm Suppose $\pi_0 \in \Xi$, $f_0\equiv 1$, and $f_1 \in \mathcal{H}$.
Further assume $D_0$ in Condition (C1) is continuous at zero and (\ref{rate}) holds. Then Condition (C3) is fulfilled.}
\end{corollary}

\begin{rem}
{\rm
{\color{red}} 
{Although $b_m$ needs to satisfy (\ref{rate}) theoretically, the rate condition is of little use in selecting $b_m$ in practice. We use a simple ad-hoc procedure that performs reasonably well in our simulations. To motivate our procedure, we let $\theta$ indicate the underlying truth of a randomly selected hypothesis from $\{H_i\}^{m}_{i=1}$. Then we have
\begin{align*}
P(\theta=0)=\frac{1}{m}\sum^{m}_{i=1}P(\theta_i=0)=\frac{1}{m}\sum^{m}_{i=1}\pi_0(i/m):=\bar{\pi}_m.       
\end{align*}
Without knowing the order information, the $p$-values follow the mixture model 
$\bar{\pi}_mf_0(x)+\left(1-\bar{\pi}_m\right)f_1(x).$
The overall null proportion $\bar{\pi}_m$ can be estimated by classical method, e.g., \cite{s2002} (in practice, we use the maximum of the two Storey's global null proportion estimates in the \texttt{qvalue} package for more conservativeness). Denote the corresponding estimator by $\hat{\pi}$. Also denote $\breve{\pi} = m^{-1}\sum^{m}_{i=1}\breve{\pi}_0(i/m)$, where $\breve{\pi}_0(i/m)  = \hat{\pi}_0(i/m) + \delta (1 - \hat{\pi}_0(i/m))$  is the calibrated null probability and $\delta$ is the amount of calibration, which is a function of $b_m$. Then it makes sense to choose $b_m \in [0, 1]$ such that the difference
$|\breve{\pi} - \hat{\pi}|$ is minimized. This results in the procedure that if the mean of $\hat{\pi}_0(i/m) $'s from the EM algorithm  (denote as $\tilde{\pi})$ is greater than the global estimate $\hat{\pi}$, $\breve{\pi}_0(i/m) =\hat{\pi}_0(i/m)$, and if the mean is less than $\hat{\pi}$, then $\breve{\pi}_0(i/m)  = \hat{\pi}_0(i/m) + \delta (1 - \hat{\pi}_0(i/m))$, where $\delta = (\hat{\pi} - \tilde{\pi}) / (1 - \tilde{\pi})$.

%In practice, we use the following simpler rule. If the mean of $\hat{\pi}_0(i/m) $'s from the EM algorithm is greater than the global estimate $\hat{\pi}$, which is taken to be the maximum of the two Storey's global null proportion estimates in the \texttt{qvalue} package, we do not adjust. If the mean is less than $\hat{\pi}$, we then calibrate the $\breve{\pi}_0(i/m) $'s to match the global estimate.
%We show in Section \ref{sec:sim} that this procedure further improves the FDR control. 
%See Figure \ref{fig15} for the results with $b_m=0.9$ without calibration.
}
}
\end{rem}

%\begin{remark}
\section{A general rejection rule}\label{sec:pval}
Given the insights from Section \ref{sec:test}, we introduce a
general rejection rule and also discuss its connection with the
recent accumulation tests in the literature, see e.g.
\cite{gwct,bc15,lb16}. Recall that our (oracle) rejection rule is
$\{\text{Lfdr}_i(x_i)\leq \lambda\}$, which can be written
equivalently as
$$\frac{f_1(x_i)}{f_0(x_i)}\geq \frac{(1-\lambda)\pi_{0i}}{\lambda(1-\pi_{0i})}.$$
Motivated by the above rejection rule, one can consider a more
general procedure as follows. Let $h(x):[0,1]\rightarrow
[0,+\infty]$ be a decreasing non-negative function such that
$\int^{1}_{0}h(x)f_0(x)dx=1$. The general rejection rule is then defined
as
\begin{align}\label{eq-rule}
h(x_i)\geq
w_i(\lambda):=\frac{(1-\lambda)\tilde{\pi}_{0i}}{\lambda(1-\tilde{\pi}_{0i})}
\end{align}
for $0\leq \lambda,\tilde{\pi}_{0i}\leq 1$. Here $h$ serves as a surrogate for the likelihood ratio $f_1/f_0.$
Set $\tilde{h}(x)=h(1-x)$. Under the
assumption that $f_0$ is symmetric about 0.5 (i.e.
$f_0(x)=f_0(1-x)$), it is easy to verify that $E[\tilde{h}(x_i)|\theta_i=0]=\int^{1}_{0}
\tilde{h}(x)f_0(x)dx=1$. We note that the false
discovery proportion (FDP) for the general rejection rule is equal
to
\begin{align*}
\text{FDP}_m(\lambda):=&\frac{\sum^{m}_{i=1}\mathbf{1}\{h(x_i)\geq
w_i(\lambda)\}(1-\theta_i)}{\sum^{m}_{i=1}\mathbf{1}\{h(x_i)\geq
w_i(\lambda)\}}
\\ =& \frac{\sum^{m}_{i=1}E[\tilde{h}(x_i)|\theta_i=0]\mathbf{1}\{h(x_i)\geq
w_i(\lambda)\}(1-\theta_i)}{\sum^{m}_{i=1}\mathbf{1}\{h(x_i)\geq
w_i(\lambda)\}}.
\end{align*}
If we set $\tilde{\pi}_{0i}=0$ for $1\leq i\leq k$ and
$\tilde{\pi}_{0i}=1$ for $k+1\leq i\leq m$, then $w_i(\lambda)=0$
for $1\leq i\leq k$ and $w_i(\lambda)=\infty$ for $k+1\leq i\leq m.$
Therefore, we have
\begin{align*}
\text{FDP}_m(\lambda)=&\frac{\sum^{k}_{i=1}E[\tilde{h}(x_i)|\theta_i=0](1-\theta_i)}{k}
\\ \approx& \frac{\sum_{i=1}^k\tilde{h}(x_i)(1-\theta_i)}{k}\leq \frac{\sum_{i=1}^k\tilde{h}(x_i)}{k},
\end{align*}
where the approximation is due to the law of large numbers. With
this choice of $\tilde{\pi}_{0i}$, one intends to follow the prior
order restriction strictly. As suggested in \cite{gwct,bc15}, a natural
choice of $k$ is given by
$$\tilde{k}=\max\left\{1\leq j\leq m: \frac{\sum_{i=1}^{j}\tilde{h}(x_i)}{j}\leq \alpha \right\},$$
which is designed to control the (asymptotic) upper bound of the
FDP.\footnote{Finite sample FDR control has been proved for this
procedure, see e.g. \cite{lb16}.} Some common choices of $\tilde{h}$
are given by
\begin{align*}
&\tilde{h}_{\text{ForwardStop}}(\lambda)=\log\left(\frac{1}{1-\lambda}\right),
\\ &\tilde{h}_{\text{SeqStep}}(\lambda)=C\mathbf{1}\{\lambda>1-1/C\},
\\ & \tilde{h}_{\text{HingeExp}}(\lambda)=C\log\left(\frac{1}{C(1-\lambda)}\right)\mathbf{1}\{\lambda>1-1/C\},
\end{align*}
for $C>0$, which correspond to the ForwardStop, SeqStep and HingeExp
procedures respectively. Note that all procedures are special cases
of the accumulation tests proposed in \cite{lb16}.

Different from the accumulation tests, we suggest to use
$\tilde{h}(x)=f_1(1-x)/f_0(1-x)$ and $\tilde{\pi}_{0i}=\pi_{0i}$.
Our procedure is conceptually sound as it is better motivated from
the Bayesian perspective, and it avoids the subjective choice of
accumulation functions.

\begin{rem}
{\rm Our setup is different from the one in \cite{gwct}, where the
authors seek for the largest cutoff $k$ so that one rejects the
first $k$ hypothesis while accepts the remaining ones. In contrast,
our procedure allows researchers to reject the $k$th hypothesis but
accept the $k-1$th hypothesis. In other words, we do not follow the
order restriction strictly. Such flexibility could result in a substantial power
increase when the order information is not very strong or even weak
as observed in our numerical studies.}
\end{rem}

Below we conduct power comparison with the accumulation tests in
\cite{lb16}, which include the ForwardStop procedure in \cite{gwct}
and the SeqStep procedure in \cite{bc15} as special cases. Let
$\tilde{h}$ be a nonnegative function with
$\int_{0}^{1}\tilde{h}(x)dx=1$ and $\nu=\int \tilde{h}(x)f_1(x)dx$.
Define
$$s_0=\max\left\{s\in [0,1]:\frac{1}{s}\int_{0}^{s}(1-\pi_0(x))dx\geq \frac{1-\alpha}{1-\nu}\right\},$$
where $s_0=0$ if the above set is empty. The asymptotic power of the
accumulation test in \cite{lb16} is given by
\begin{align*}
\text{Power}_{\text{AT}}=\frac{\int_{0}^{s_0}(1-\pi_0(x))dx}{\int_{0}^{1}(1-\pi_0(x))dx}.
\end{align*}
%\textbf{Question: How to compare $\text{Power}$ with $\text{Power}_{AT}$? From the above derivation, I feel that our procedure should enjoy certain optimality. We may want to present some numerical results regarding the %asymptotic power comparison here.}
Notice that the accumulation test rejects the first $s_0$
hypotheses in the ordered list, which is equivalent to setting the
threshold $s_i=1$ for $1\leq i\leq s_0$ and $s_i=0$ otherwise.
Suppose $s_0$ satisfies %that
$$\frac{1}{s_0}\int_{0}^{s_0}(1-\pi_0(x))dx=\frac{1-\alpha}{1-\nu}.$$
Then after some re-arrangements, we have
$$\frac{\int_{0}^{s_0}\pi_0(x)dx}{s_0}\leq \frac{\int_{0}^{s_0}\pi_0(x)dx}{s_0}+\frac{\nu\int_{0}^{s_0}(1-\pi_0(x))dx}{s_0}=\alpha,$$
which suggests that the accumulation test controls the mFDR defined
in (\ref{mfdr}) at level $\alpha.$ By the discussion in Section
\ref{sec:setup}, the optimal thresholds are the level surfaces of
the Lfdr. Therefore the proposed procedure is more powerful than the
accumulation test asymptotically.
%\end{remark}

\subsection{Asymptotic power analysis}\label{sec:power}
We provide asymptotic power analysis for the proposed method.
%Following
%the setup in Section \ref{sec:iso}, assume that
%$\pi_{0i}=\pi_0(i/m)$ for some non-decreasing function $\pi_0$.
%Define $g(x)=\sup\{t\in [0,1]: f_1(t)/f_0(t)\geq x\}.$ Denote by $F_i$ the distribution function of $f_i$ for $i=0,1.$
%Note that
%for large $m$, the realized power of the Lfdr-based procedure %is
%approximately equals to
%\begin{align*}
%\text{Power}_{\text{Lfdr}}=&\frac{\sum_{i=1}^{m}\mathbf{1}\{\theta_i=1,\text{Lfdr}_i(x_i)\leq
%\lambda_0\}}{\sum_{i=1}^{m}\mathbf{1}\{\theta_i=1\}}
%\\ \approx& \frac{m^{-1}\sum_{i=1}^{m}P(\text{Lfdr}_i(x_i)\leq %\lambda_0|\theta_i=1)P(\theta_i=1)}{m^{-1}\sum_{i=1}^{m}P(\theta_i=1)}
%\\ \approx& \frac{\int_{0}^{1}(1-\pi_0(x))F_1\left(g\circ %w(\lambda_0,x)\right)dx}{\int^{1}_{0}(1-\pi_0(x))dx}
%\end{align*}
%where ``$\circ$" denotes the composition of two functions,
%$w(\lambda,x)=\frac{\pi_0(x)(1-\lambda)}{(1-\pi_{0}(x))\lambda}$
%and the first approximation is due to the law of large numbers. Here
%$\lambda_0$ is the largest $\lambda \in [0,1]$ such that
%$$\frac{\int^{1}_{0}\pi_0(x)F_0(g\circ w(\lambda,x))dx}{\int^{1}_{0}\left\{\pi_0(x)F_0(g\circ %w(\lambda,x))+(1-\pi_0(x))F_1(g\circ w(\lambda,x))\right\}dx}\leq \alpha.$$
In particular, we have the following result concerning the asymptotic power of the Lfdr procedure in Section
\ref{sec:asym}.
\begin{theorem}\label{thm:power}
Suppose Conditions (C1)-(C3) hold and additionally assume that
\begin{align*}
&\frac{1}{m}\sum^{m}_{i=1}\mathbf{1}\{\theta_i=0\}\rightarrow \kappa_0,\\
&\frac{1}{m}\sum_{i=1}^{m}\mathbf{1}\{\theta_i=1,\text{Lfdr}_i(x_i)\leq
\lambda\}\rightarrow^p D_2(\lambda),
\end{align*}
for a continuous function $D_2$ of $\lambda$ on [0,1]. Let $\lambda_0$ be the largest $\lambda \in [0,1]$ such that $R(\lambda)\leq \alpha$ and for any small enough $\epsilon$, $R(\lambda_0-\epsilon)<\alpha.$ Then we have
\begin{align*}
\text{Power}_{\text{Lfdr}}&:=\frac{\sum_{i=1}^{m}\mathbf{1}\{\theta_i=1,\widehat{\text{Lfdr}}_i(x_i)\leq
\hat{\lambda}_m\}}{\sum_{i=1}^{m}\mathbf{1}\{\theta_i=1\}\vee 1}
\rightarrow^p \frac{D_2(\lambda_0)}{1-\kappa_0}.
\end{align*}
\end{theorem}
Recall that in Section \ref{sec:setup}, we have shown that the step-up procedure has the highest expected number of true positives amongst all $\alpha$-level FDR rules. This result thus sheds some light on the asymptotic optimal power amongst all $\alpha$-level FDR rules when the number of hypothesis tests goes to infinity.

\begin{rem}
{\rm Under the two-group mixtue model (\ref{prior-prob})-(\ref{mix-model}) with $\pi_{0i}=\pi_0(i/m)$ for some non-decreasing function $\pi_0$, we have
$m^{-1}\sum^{m}_{i=1}P(\theta_i=0)=m^{-1}\sum^{m}_{i=1}\pi_0(i/m)\rightarrow \int^{1}_{0}\pi_0(x)dx$ as monotonic functions are Riemann integrable. Thus $\kappa_0=\int^{1}_{0}\pi_0(x)dx$. Define $g(x)=\sup\{t\in [0,1]: f_1(t)/f_0(t)\geq x\}$
and $w(\lambda,x)=\frac{\pi_0(x)(1-\lambda)}{(1-\pi_{0}(x))\lambda}$. Denote by $F_1$ the distribution function of $f_1$. Then we have
\begin{align*}
\frac{1}{m}\sum_{i=1}^{m}P(\theta_i=1,\text{Lfdr}_i(x_i)\leq
\lambda)=&\frac{1}{m}\sum_{i=1}^{m}P(\theta_i=1)P(\text{Lfdr}_i(x_i)\leq
\lambda|\theta_i=1)
\\=&\frac{1}{m}\sum_{i=1}^{m}(1-\pi_0(i/m))F_1\circ g\circ w(\lambda,i/m)
\\ \rightarrow& \int^{1}_{0}(1-\pi_0(x))F_1\circ g\circ w(\lambda,x)dx,
\end{align*}
where ``$\circ$" denotes the composition of two functions, and we have used the fact that $F_1\circ g\circ w$ is monotonic and thus Riemann integrable. So $D_2(\lambda)=\int^{1}_{0}(1-\pi_0(x))F_1\circ g\circ w(\lambda,x)dx.$
}
\end{rem}

%As an illustration, we consider
%the following simple example.
%\begin{example}
%{\rm Following the setup in Example \ref{exam1}, we assume that
%$$\pi_0(p)=\frac{\pi}{\pi+(1-\pi)f_1(p)},$$ $f_0(p)=1$ and
%$F_1(p)=\text{Beta}(p;a,b)$, where $\text{Beta}(\cdot;a,b)$ denotes
%the beta distribution function with parameters $(a,b)$. We set
%$(a,b)=(1,40)$ and the nominal level $\alpha=0.2.$ As seen from the
%right panel of Figure \ref{fig:p2}, the LFDR and the hybrid
%procedures are clearly more powerful than the ForwardStop and the
%HingeExp procedures in Li and Barber (2016). }
%\end{example}

%\begin{figure}[ht!]
%\centering
%\includegraphics[height=7cm,width=7cm]{beta-density.pdf}
%\includegraphics[height=7cm,width=7cm]{power.pdf}
%\caption{Left panel: density curves for the beta distribution with
%$(a,b)=(1,40)$; Right panel: Asymptotic powers for the accumulation
%tests, the LFDR-based procedure and the hybrid procedure. The
%nominal FDR level is 20\%.}\label{fig:p2}
%\end{figure}

\section{Two extensions}\label{sec:extend}
\subsection{Grouped hypotheses with ordering}
Our idea can be extended to the case where the hypotheses can be
divided into $d\geq 2$ groups within which there is no explicit
ordering but between which there is an ordering. One can simply
modify (\ref{em-1}) by considering the problem,
\begin{equation}\label{group-order}
(\hat{\pi}_{01}^{(t+1)},\dots,\hat{\pi}_{0d}^{(t+1)})=\argmin
\sum_{j=1}^{m}\left\{\tilde{Q}_{j}^{(t)}-\pi_{s(j)}\right\}^2\footnote{This
optimization problem can be solved by slightly modifying the PAVA by
averaging the estimators within each group.},
\end{equation}
subject to $0\leq \pi_{1}\leq \cdots \leq \pi_{d}\leq 1$, where
$s(j)\in\{1,2,\dots,d\}$ is the group index for the $j$th
hypothesis. A particular example is about using the sign to improve
power while controlling the FDR. Consider a two-sided test where the
null distribution is symmetric and the test statistic is the
absolute value of the symmetric statistic. The sign of the statistic
is independent of the $p$-value under the null. If we have {\it a priori}
belief that among the alternatives, more hypotheses have true
positive effect sizes than negative ones or vice versa, then sign
could be used to divide the hypotheses into two groups such that
$\pi_1\leq \pi_2$ (or $\pi_1\geq \pi_2$).

\subsection{Varying alternative distributions}
In model (\ref{prior-prob}), we assume that the success probabilities
$\pi_{0i}, i =1, \ldots, m$ vary with $i$ while $F_1$ is independent of $i$. This
assumption is reasonable in some applications but it can be
restrictive in other cases. We illustrate this point via a simple
example described below.
\begin{example}
{\rm For $1\leq i\leq m$, let $\{x_{ik}\}_{k=1}^{n_i}$ be $n_i$
observations generated independently from $N(\mu_i,1)$. Consider the
one sided $z$-test $Z_i=\sqrt{n_i}\bar{x}_i$ with
$\bar{x}_i=n_i^{-1}\sum_{k=1}^{n_i}x_{ik}$ for testing
$$H_{i0}:\mu_i=0\quad \text{vs}\quad H_{ia}:\mu_i<0.$$
The $p$-value is equal to $p_i=\Phi(\sqrt{n_i}\bar{x}_i)$ and the $p$-value distribution under the alternative hypothesis is given by
$$F_{1i}(x)=\Phi\left(\Phi^{-1}(x)-\sqrt{n_i}\mu_i\right),$$
with the density
$$f_{1i}(x)=\frac{\phi(\Phi^{-1}(x)-\sqrt{n_i}\mu_i)}{\phi(\Phi^{-1}(x))}=\exp\left(\frac{2\sqrt{n_i}\mu_i\Phi^{-1}(x)-n_i\mu^2_i}{2}\right).$$
By prioritizing the hypotheses based on the values of
$\sqrt{n_i}\mu_i$, one can expect more discoveries. Suppose
$$n_1\mu_1^2\leq n_2\mu_2^2\leq \dots\leq n_m\mu_m^2.\footnote{This is the case if $\mu_i=\mu$ and $n_1\leq n_2\leq \cdots \leq n_m.$}$$
One can consider the following problem to estimate $\pi$ and $\mu_i$
simultaneously,
$$\argmax_{\pi\in [0,1],r_m\leq r_{m-1}\leq \cdots\leq r_1<0}\sum_{i=1}^{m}\log\left\{\pi+(1-\pi)\exp\left(\frac{2r_i\Phi^{-1}(p_i)-r^2_i}{2}\right)\right\}.$$
This problem can again be solved using the EM algorithm together
with the PAVA.}
\end{example}

Generally, if the $p$-value distribution under the alternative hypothesis, denoted by $F_{1i}$, is
allowed to vary with $i$, model (\ref{prior-prob})-(\ref{mix-model})
is not estimable without extra structural assumptions as we only have
one observation that is informative about $F_{1i}$. On the other
hand, if we assume that $F_{1i}:=F_{1,i/m}$ which varies smoothly
over $i$, then one can use non-parametric approach to estimate each
$F_{1,i/m}$ based on the observations in a neighborhood of $i/m$.
%The algorithm can be integrated into our general EM-algorithm presented in Section
%\ref{general:alg}. Specially, in the M-step, we estimate the density
%function of $F_{1i}$ by
%\begin{align}\label{obj-shape} f_{1i}^{(t+1)}=\argmax_{f_1}
%\sum^{m}_{j=1}K_h(x_j-x_i)(1-Q_{j}^{(t)})\log f_1(x_j),
%\end{align}
%where $K_h(\cdot)=K(\cdot/h)/h$ with $K$ being a kernel function and $h$ being a bandwidth parameter.
However, this method requires the estimation of $m$ density
functions at each iteration, which is computationally expensive for
large $m$. To reduce the computational cost, one can divide the
indices into $K$ consecutive bins, say $S_1,S_2,\dots,S_K,$ and
assume that the density remains unchanged within each bin. In the
M-step, we update $f_{1i}$ via
\begin{align}\label{obj-shape2}
f_{1i}^{(t+1)}=\argmax_{\tilde{f}_1\in\mathcal{H}} \sum_{j\in S_i}(1-Q_{j}^{(t)})\log
\tilde{f}_1(x_j),
\end{align}
for $i=1,2,\dots,K.$ For small $K$, the computation is relatively efficient. We note that this strategy is related to the independent hypothesis weighting proposed in \cite{Ignatiadis2016,Ignatiadis2017}, which divides the p-values into several bins and estimate the cumulative distribution function (CDF) of the p-values in each stratum. Our method is different from theirs in the following aspect: the estimated densities will be used in constructing the optimal rejection rule, while in their procedure, the varying CDF is used as an intermediate quantity to determine the thresholds for p-values in each stratum. In other words, the estimated CDFs are not utilized optimally in constructing the rejection rule.

%\textbf{Maybe we can present some numerical results for these two modified algorithms.}

\section{Simulation studies}\label{sec:sim}
\subsection{Simulation setup}\label{simset}
\begin{figure}
\caption[Simulation Strategy]{Simulation Strategy. (a) The
distribution of probabilities of being null ($\pi_{0i}, i =1, \ldots, m$) for three scenarios
representing weakly, moderately and highly informative
auxiliary information (from bottom to top). Different levels of signal
density are simulated. (b) Distribution of the realized
$\pi_{0i}$ for alternatives and nulls from one
simulated dataset.} \label{fig:sim:0}

\centering
\includegraphics[scale=0.42]{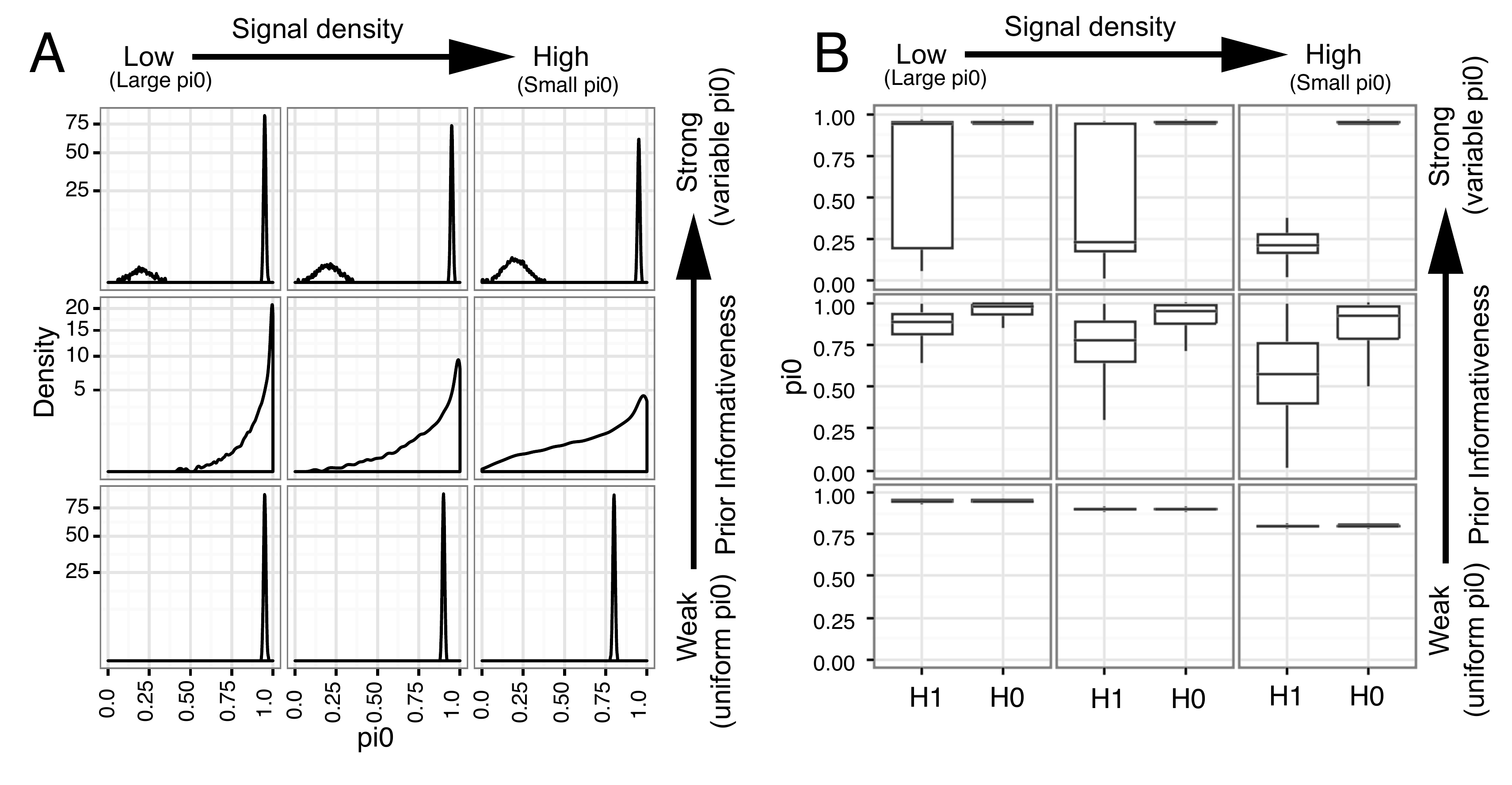}
 %by adjusting the parameter of the
%distribution.
\end{figure}

We conduct comprehensive simulations to evaluate the finite-sample
performance of the proposed method and compare it to competing
methods. For simplicity,  we directly simulate $z$-values for
$m{=}10,000$ hypotheses. All simulations are replicated 100 times except for the global null, where the results are based on 2,000 Monte Carlo replicates. 
We simulate different combinations of
signal density (the percentage of alternative) and signal strength
(the effect size of alternative) since these are two main factors
affecting the power of multiple testing procedures.   We first generate the hypothesis-specific null
probability ($\pi_{0i}$), upon which the truth, i.e., null or alternative, is simulated. Afterwards, we generate $z$-values based on the truth of the hypothesis.    %To  demonstrate the
%power improvement using prior information,
We first use  $\pi_{0i}$ as the auxiliary covariate. Later, we will study the effect of using noisy $\pi_{0i}$ as auxiliary covariate.
Three scenarios, representing weakly, moderately and highly
informative auxiliary information, are simulated based on the distribution of $\pi_{0i}$ (Figure \ref{fig:sim:0}(a)), where  the informativeness of the auxiliary covariate is determined based on its ability to separate alternatives from nulls (Figure \ref{fig:sim:0}(b)).
In the weakly informative scenario, we make $\pi_{0i}$'s similar for all hypotheses by
simulating  $\pi_{0i}$'s from a highly concentrated normal
distribution (truncated on the unit interval $[0,1]$)
$$\pi_{0i} \sim N_C(\mu_w, 0.005^2). $$
 In the moderately informative scenario, we allow $\pi_{0i}$ to vary across hypotheses with moderate variability. This is achieved by simulating $\pi_{0i}$'s from a beta distribution
 $$\pi_{0i}  \sim \text{Beta}(a, b).$$  In the highly informative scenario, $\pi_{0i}$'s are simulated from a mixture of a truncated normal  and a highly concentrated truncated normal distribution
 $$\pi_{0i} \sim \pi_hN_C(\mu_{h1}, \sigma_{h1}^2) + (1 - \pi_h) N_C(\mu_{h2}, 0.005^2),$$
 %that represents a clear separation --- one mixture is for the null group and the other mixture is for the non-null group.
  which represents two groups of hypotheses with strikingly different probabilities of being null.
% This corresponds to the real-world scenario, where we know a small number of hypotheses are very likely to be true. % but we do not have good knowledge for the rest majority.
Since the expected alternative proportion is  \newline $\sum_{i=1}^m{(1 - \pi_{0i})} / m$,  we adjust the parameters $\mu_w, a, b, \pi_h, \mu_{h1},$ $\sigma_{h1}^2$ and $\mu_{h2}$ to
 achieve approximately 5\%, 10\% and 20\% signal density level.  Figure \ref{fig:sim:0}(a) shows the distribution of $\pi_{0i}$ for the three scenarios. Based on $\pi_{0i}$,
 the underlying truth $\theta_i$ is simulated from $$\theta_i \sim \text{Bernoulli}(1 - \pi_{0i}).$$ Figure \ref{fig:sim:0}(b) displays the distribution of
 $\pi_{0i}$ for  $\theta_i{=1}$ and $\theta_i{=0}$ from one simulated dataset. As the difference in $\pi_{0i}$ between $H_1$ and $H_0$ gets larger,
the auxiliary covariate becomes more informative. % in distinguishing $H_1$ from $H_0.$ %, %as we move from weakly to strongly informative prior.
 Finally, we simulate independent $z$-values using $$z_i \sim  N(k_s\theta_i, 1), $$ where $k_s$ controls the signal strength and $k_s{=} 2, 2.5$ and  $3$ are chosen to
 represent  weak, moderate and strong signal, respectively.  We convert $z$-values to $p$-values using the formula $p_i = 1 - \Phi(z_i)$. The proposed method accepts $p$-values and
 $\pi_{0i}$s as input. The specific parameter values mentioned above could be found in \url{https://github.com/jchen1981/OrderShapeEM}.

 %are used as the input for the proposed method.

 To examine the robustness of the proposed method, we vary the simulation setting in different ways. Specifically, we investigate:

 \begin{itemize}
 \item[1.] {\it Skewed alternative distribution}. Instead of simulating normal $z$-values for the alternative group, we simulate $z$-values from a non-central gamma distribution with the shape parameter $k{=2}$. The scale and non-centrality parameters of the non-central gamma distribution are chosen to match the mean and variance of the normal distribution for the alternative group under the basic setting.
 \item[2.] {\it Correlated hypotheses}. %The theory of the proposed method assumes  independence between the hypotheses.  The assumption may not hold for real data sets.
 Our theory allows certain forms of dependence. We then simulate correlated $z$-values, which are drawn from a multivariate normal distribution with a block correlation structure. The order of $\pi_{0i}$ is random with respect to the block structure. Specifically, we divide the $10,000$ hypotheses into $100$ blocks and each block is further divided into two sub-blocks of equal size.  Within each sub-block, there is a constant positive correlation ($\rho{=}0.5$). Between the sub-blocks in the same block, there is a constant negative correlation ($\rho{=}{-}0.5$). Hypotheses in different blocks are independent. We use $p = 8$ to illustrate. The correlation matrix is
 $$\begin{pmatrix}
1 & 0.5 &  0.5 & 0.5 & -0.5 & -0.5 &  -0.5 & -0.5 \\
0.5 & 1 &  0.5 & 0.5 & -0.5 & -0.5 &  -0.5 & -0.5 \\
0.5 & 0.5 & 1 & 0.5 & -0.5 & -0.5 &  -0.5 & -0.5 \\
0.5 & 0.5  & 0.5 & 1 & -0.5 & -0.5 &  -0.5 & -0.5 \\
-0.5 & -0.5  & -0.5 & -0.5 & 1 & 0.5 &  0.5 & 0.5 \\
-0.5 & -0.5 & -0.5 & -0.5 & 0.5 & 1 &  0.5 & 0.5 \\
-0.5 & -0.5  & -0.5 & -0.5 & 0.5 & 0.5 &  1 & 0.5 \\
-0.5 & -0.5  & -0.5 & -0.5 & 0.5 & 0.5 &  0.5 & 1 \\
\end{pmatrix}.$$
  \item[3.] {\it Noisy auxiliary information}. In practice,  the auxiliary data can be very noisy.  To examine the effect of noisy auxiliary information, we shuffle half or all the $\pi_{0i}$, representing moderately and completely noisy order.
 %from the ``highly informative prior" scenario by shuffling half or all $\pi_{0i}s$.

 \item[4.] {\it A smaller number of alternative hypotheses and a global null}.  It is interesting to study the robustness of the proposed method under an even more sparse signal. We thus simulate 1\% alternatives out of 10,000 features. 
 % {\color{blue}\{Results are in Normal\_Power\_grid\_bar\_1\%.pdf and Normal\_FDR\_grid\_bar\_1\%.pdf\}.}  
  We also study the error control under a global null, where all the hypotheses are nulls. Under the global null, We increased the number of Monte Carlo simulations to 2,000 times to have a more accurate estimate of the FDR.
  % {\color{blue} \{Results are in Normal\_Power\_grid\_bar\_0\%.pdf and Normal\_FDR\_grid\_bar\_0\%.pdf\}}. 
 \item[5.] {\it Varying $f_1$ across alternative hypotheses}. We consider the case where among the alternative hypotheses, the most promising 20\% hypotheses (i.e., those with the lowest prior order) follow $\text{Unif}(0, 0.02)$ and the remaining p-values are derived from the z-values (see the setting of Figure 2). 
 %{\color{blue}\{Results are in Normal\_Power\_grid\_bar\_varyingf1\%.pdf and Normal\_FDR\_grid\_bar\_varyingf1\%.pdf. The conclusion is the same.\}.} 
  \item[6.] {\it Varying $f_0$ across null hypotheses}. Similar to the case of varying $f_1$, we sample the p-values of 20\% of the null hypotheses with the highest prior order from $\text{Unif}(0.5, 1)$, which mimics the composite null situations. The remaining p-values are derived from the z-values as above.
  %{\color{blue} \{Results are in Normal\_Power\_grid\_bar\_varyingf0\_s1\%.pdf and Normal\_FDR\_grid\_bar\_varyingf0\_s1\%.pdf.  We see only AdaPT can control FDR, which is expected since AdaPT assumes the null p-value distribution to be symmetrical or mirror-conservative. AdaptiveSeqStep also works well under this scenario \}.} 
 %In the second scenario, we add $\mu^z_i \sim \text{Unif}(0, 0.5)$ to $z_i$ for the nulls, which leads to a right-skewed null p-value distribution with a spike at the left end. 
 % {\color{blue}\{Results are in Normal\_Power\_grid\_bar\_varyingf0\_s2\%.pdf and Normal\_FDR\_grid\_bar\_varyingf0\_s2\%.pdf.  We can see none of the compared methods can control the FDR. The take-home message is that addressing varying null hypotheses (composite nulls) are challenging without certain assumption of specific forms of $f_0$ density.  It will be an interesting future research topic \}.  }
 
  \end{itemize}

We compare the proposed method (OrderShapeEM) with classical multiple testing methods that do not utilize external covariates (BH and ST) and recent multiple testing procedures that exploit auxiliary information (AdaPT, SABHA, AdaptiveSeqStep). Detailed descriptions of these methods are provided in the appendix. The FDP estimate of AdaPT involves a  finite-sample correction term +1 in the numerator. The +1 term yields a conservative procedure and could lose power when the signal density is low. To study the effect of the correction term, we also compared to AdaPT+, where we removed the correction term +1 in the numerator. However, we observed a significant FDR inflation when the signal density is low, see Figure \ref{fig14} in the Appendix. 
%{\color{blue}\{See Normal\_FDR\_grid\_bar\_AdaPT+1.pdf for example.\} } 
We thus compared to  AdaPT procedure with correction term throughout the simulations.

\subsection{Simulation results}
We first discuss the simulation results of \textit{Normal alternative distribution}.
%Results based on \textit{Skewed non-null distribution} and \textit{Noisy auxiliary information} have similar patterns and are relegated in the Appendix, see Figures \ref{fig6}-\ref{fig7}.
%
%In Figure \ref{fig3}, we plot FDR control and power comparison with different methods when $z$ values under the null hypothesis follow $N(0,1)$ and $z$ values under the alternative hypothesis follow a non-central gamma distribution. In Figure \ref{fig3}(a), the dashed line indicates the pre-specified FDR control level $0.05$ and the error bars represent empirical $95\%$ confident interval. We observe that the proposed method controls FDR in most scenarios except for weak signal, weak to moderate auxiliary information and sparse signal density. SABHA, ST and OrdinaryEM all control FDR reasonably well. AS is conservative most of the time except for density signal density and moderate to strong auxiliary information. BH procedure is conservative as excepted, especially when signal density is dense. In Figure \ref{fig3}(b), we plot compar
\begin{figure}\label{fig3-1}
%\caption{Performance under normal alternative distribution. }\label{fig3-1}
%\hfill
%\begin{subfigure}[t]%{0.5\textwidth} %[c] %{0.98\textwidth}
\centering
\caption{FDR control }\label{fig3-1}
\includegraphics[scale=0.8]{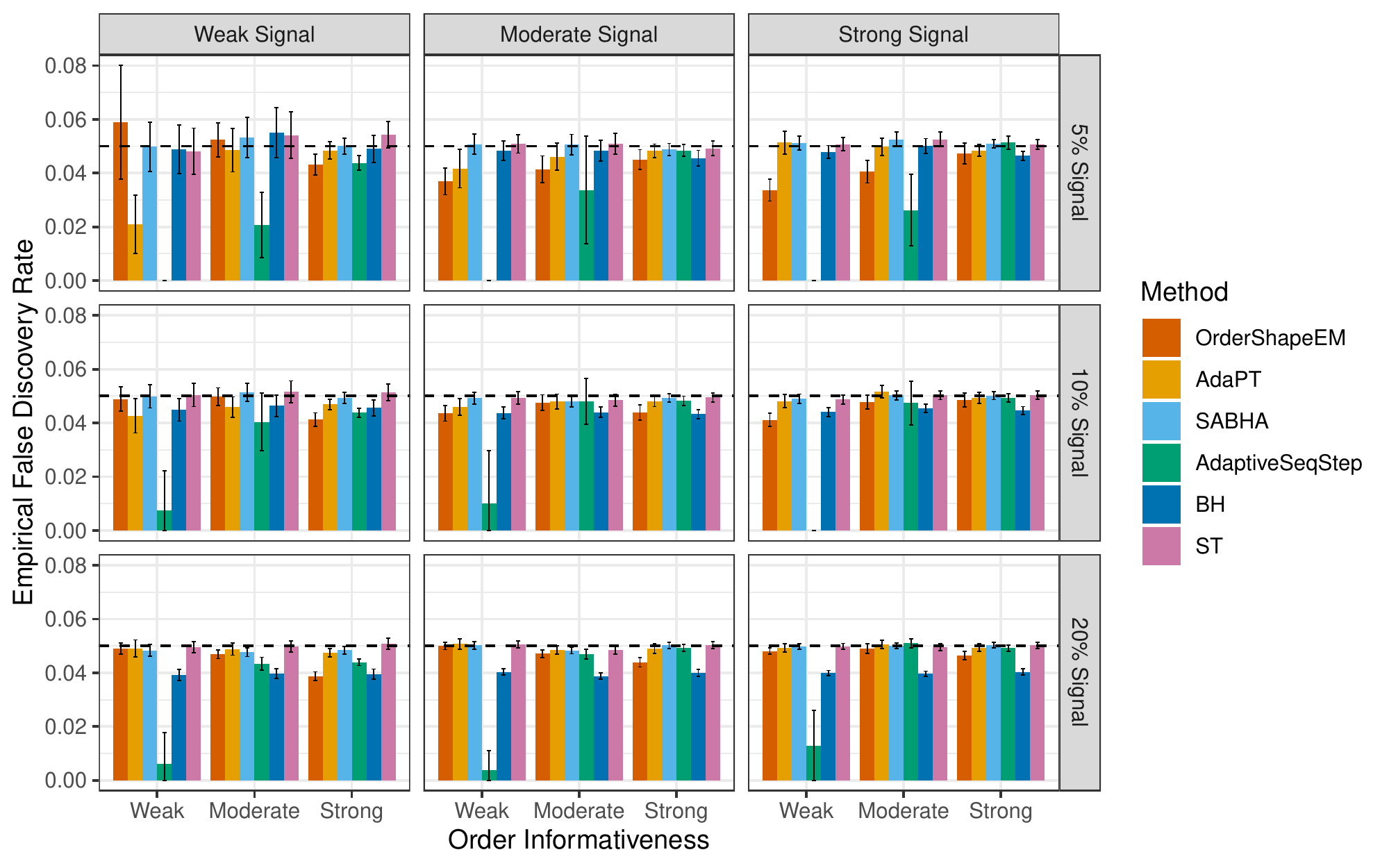}
%\end{subfigure}
%\hfill
%\begin{subfigure}[t]%{0.5\textwidth} %[c]%{0.98\textwidth}
%\centering
%\caption{power comparison }
%\includegraphics[scale=0.5]{Figure2b.pdf}
%\end{subfigure}
%\caption[Comparison of FDR control when the alternative distribution
%is gamma dist.]{Comparison of FDR control under skewed non-null
%distribution at different levels of signal strength (columns),
%signal density (rows) and order informativeness (within each box).
%Z-scores are independent and normally distributed for nulls but have
%a skew distribution for non-nulls. Bar plots show the average false
%discovery proportions  across 100 simulation runs. Error bars
%represent the normal-based 95\% CIs.The dashed line indicates the
%target level.} \label{fig:sim:3}
\end{figure}

\begin{figure}%\label{fig3-2}
\caption{power comparison }\label{fig3-2}
\includegraphics[scale=0.8]{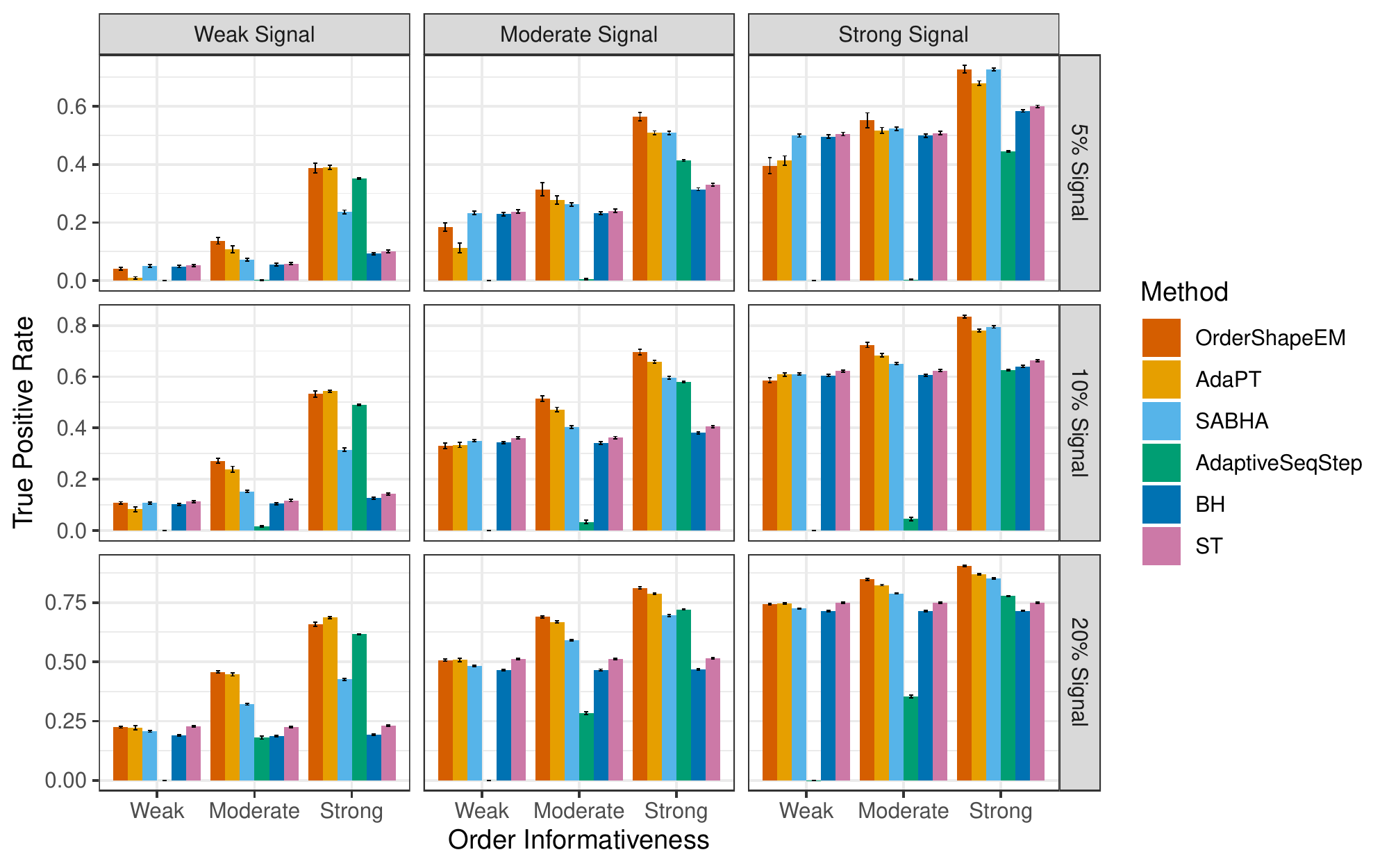}
\end{figure}
%
%\begin{figure}
%\centering
%\includegraphics[scale=0.75]{Gamma_Power_grid_bar}
%\caption[Comparison of power when the alternative distribution is
%gamma dist.]{Comparison of power under skewed non-null distribution
%at different levels of signal strength (columns), signal density
%(rows) and order informativeness (within each box).  Z-scores are
%independent and normally distributed for nulls but have a skew
%distribution for non-nulls. Bar plots show the average true positive
%rates  across 100 simulation runs. Error bars represent the
%normal-based 95\% CIs.} \label{fig:sim:4}
%\end{figure}
{\color{red}}{ In Figure \ref{fig3-1} and \ref{fig3-2}, we present FDR control and power comparison with different methods when $z$-values under the null hypothesis follow $N(0,1)$ and $z$-values under the alternative hypothesis follow a normal distribution. In Figure \ref{fig3-1}, the dashed line indicates the pre-specified FDR control level $0.05$ and the error bars represent empirical $95\%$ confidence intervals. We observe that all procedures
%controls FDR in most scenarios except for weak signal, weak to moderate auxiliary information and sparse signal density.
control the FDR sufficiently well across settings and no FDR inflation has been observed. Adaptive SeqStep is conservative most of the time especially when the signal is sparse and the auxiliary information is weak or moderate. AdaPT is conservative under sparse signal and weak auxiliary information.  The proposed procedure OrderShapeEM generally controls the FDR at the target level with some conservativeness under some settings. As expected, ST procedure controls the FDR at the target level while BH procedure is more conservative under dense signal.  In Figure \ref{fig3-2}, we observe that OrderShapeEM is overall the most powerful when the auxiliary information is not weak.  When the auxiliary information is weak and the signal is sparse,  OrderShapeEM could be less powerful than BH/ST. Close competitors are AdaPT and SABHA. However, AdaPT is significantly less powerful when the signal is sparse and the auxiliary information is weak.  AdaPT is also computationally more intensive than the other methods. SABHA performs well when the signal is strong but becomes much less powerful than OrderShapeEM and AdaPT as the signal weakens.  Adaptive SeqStep has good power for dense signal and moderate to strong auxiliary information. However, it is powerless when auxiliary information is weak. If auxiliary information is weak, SABHA,  ST and BH have similar power, while Adaptive SeqStep has little power. Under this scenario, incorporating auxiliary information does not help much. All methods become more powerful with the increase of signal density and signal strength.}

\section{Data Analysis}\label{sec:data}
We illustrate the application of our method by analyzing data from publicly available genome-wide association studies (GWAS). We use datasets from two large-scale GWAS of coronary artery disease (CAD) in different populations (CARDIoGRAM and C4D).  CARDIoGRAM is a meta-analysis of $14$ CAD genome-wide association studies, comprising $22,233$ cases and $64,762$ controls of European descent \citep{Schunkert11}. The study includes $2.3$ million single nucleotide polymorphisms (SNP). In each of the $14$ studies and for each SNP, a logistic regression of CAD status was performed on the number of copies of one allele, along with suitable controlling covariates. C4D is a meta-analysis of $5$ heart disease genome-wide association studies, totaling $15,420$ CAD cases and $15,062$ controls \citep{CAD11}. The samples did not overlap those from CARDIoGRAM. The analysis steps were similar to CARDIoGRAM. A total of $514,178$ common SNPs were tested in both the CARDIoGRAM and C4D association analyses. Dataset can be downloaded from \url{http://www.cardiogramplusc4d.org}. Available data comprise of a bivariate $p$-value sequence $(x_{1i}, x_{2i}),$ where $x_{1i}$ represents $p$-values from the CARDIoGRAM dataset and $x_{2i}$ represents $p$-values from the C4D dataset, $i =1, \ldots, 514,178.$

We are interested in identifying SNPs that are associated with CAD. Due to the shared genetic polymorphisms between populations, information contained in $x_{i1}$ can be helpful in the association analysis of $x_{2i}$ and vice versa. We thus performed two separate analyses, where we conducted FDR control on $x_{1i}$ and $x_{2i}$ respectively,  using  $x_{2i}$ and $x_{i1}$ as the auxiliary covariate. %Intuitively, if $p_{1i}$ is small for the $i$th SNP, most likely, the probability that the $i$th SNP is associated with CAD ($\pi_{0i}$) is low. Thus, information contained in $p_{i1}$ can be helpful in the association analysis of $p_{2i}$ and vice versa.

\begin{figure}
\caption{Comparison of the number of discoveries at different pre-specified FDR level (left panels) as well as the estimates of $\pi_0$ (middle panels) and $f_1$ (right panels).} (a) Analysis of C4D data with CARDIoGRAM data as auxiliary information; (b) Analysis of CARDIoGRAM data with C4D data as auxiliary information.\label{fig:real:3}
\centering
\includegraphics[scale=0.35]{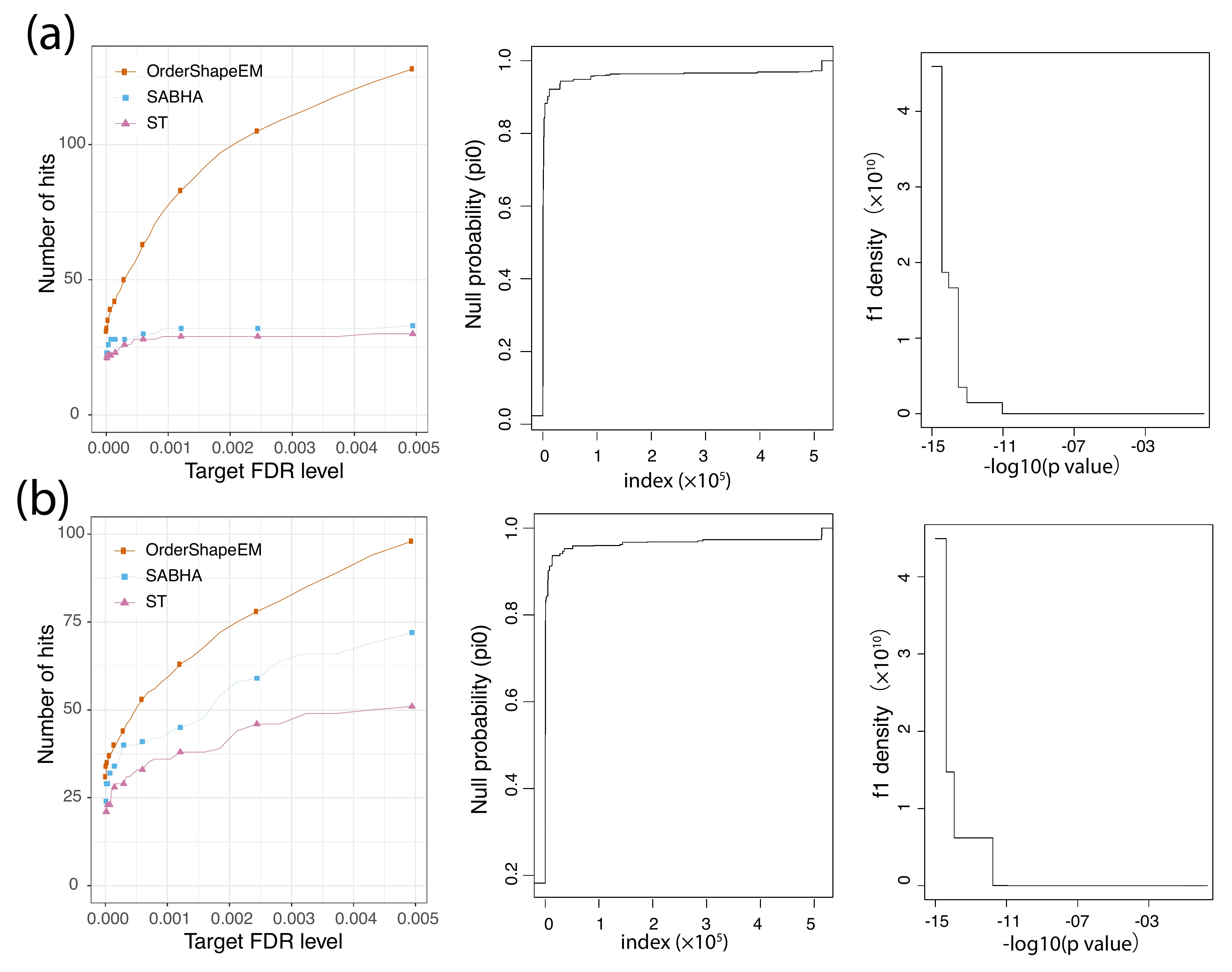}
%\caption[Comparison of the number of discoveries  at different
%target levels for the GWAS data.]{Comparison of the number of discoveries  at different
%target levels for the two GWAS data sets. A. FDR control on C4D data with  CARDIoGRAM data as auxiliary.
%B. FDR control on  CARDIoGRAM data with  C4D data as auxiliary. For each method, the plot shows the number of
%discoveries made (i.e. the number of SNPs selected as showing
%significant association with coronary artery disease), at a range of
%target FDR values.} \label{fig:real:3}
\end{figure}

In the analysis, we compare the proposed OrderShapeEM, robust method that incorporates auxiliary information (SABHA) and method that does not incorporate auxiliary information (ST). As BH was outperformed by ST and Adaptive SeqStep by SABHA, we only included ST and SABHA in the comparison. AdaPT was not able to complete the analysis within 24 hours and was not included either. The results are summarized in Figure \ref{fig:real:3}. From Figure \ref{fig:real:3}(a), we observe that at the same FDR level, the proposed OrderShapeEM made significantly more discoveries than SABHA and ST.  SABHA procedure, which incorporates the auxiliary information, picked up more SNPs than the ST procedure. %In contrast, AdaptiveSeqStep is not able to make any discoveries, possibly due to a moderately informative prior ordering.
The performance of OrderShapeEM is consistent with the weak signal scenario, where a significant increase in power has been observed (Figure \ref{fig3}(b)). Due to disease heterogeneity, signals in the genetic association studies are usually very weak. Thus, it can be extremely helpful to incorporate auxiliary information to improve power. The power difference becomes even larger at higher target FDR level. Figure \ref{fig:real:3}(b) shows similar patterns.

%
%We finally tested our method on a GWAS data set for the study of the
%single nucleotide  polymorphisms (SNPs) associated with Coronary Artery Disease (CAD). The data
%were from two independent, publicly available GWAS meta-analyses of
%coronary artery disease available from the CARDIoGRAMplusC4D
%Consortium (http://www.cardiogramplusc4d.org/downloads/)
%\cite{Schunkert11, CAD11}.  The larger CARDIoGRAM study  consists of
%22,233 cases and  64,762 controls, and the C4D data consists of
%15,420 cases and 15,062 controls. A total of 514,178 common SNPs
%were  tested in both the C4D and CARDIoGRAM association analyses.
%In this example, we  first used the p-values from CARDIoGRAM study to
%create prior ordering and performed the analysis on the C4D dataset (Figure \ref{fig:real:3}A).
%%Compared to gene expression and DNA methylation datasets,
%Due to disease heterogeneity,  genetic association signals are usually very weak and a large number of
%samples are required to ensure adequate statistical power to detect
%associations.  Consistent with the performance under weak-signal
%scenarios, OrderShapeEM was far more powerful than the other
%competing methods including SABHA.  The power difference became even
%larger when a higher target FDR level was used, indicating the
%effectiveness of using prior ordering information to boost power for
%weaker signals.  The same pattern was observed for OrderShapeEM on the CARDIoGRAM dataset using C4D dataset as axillary information (Figure \ref{fig:real:3}B).

\begin{figure}
\caption[Overlap of associated SNPs with FDR $<$ 0.001.]{ Venn diagram showing the overlap of significant SNPs (FDR $<$ 0.001) between methods using or not using auxiliary information. Left to right: ST procedure on C4D data; OrderShape EM on C4D data with CARDIoGRAM data as auxiliary; OrderShapeEM on CARDIoGRAM data with  C4D data as auxiliary; and ST procedure on CARDIoGRAM data.} \label{fig:real:4}
\centering
\includegraphics[scale=0.5]{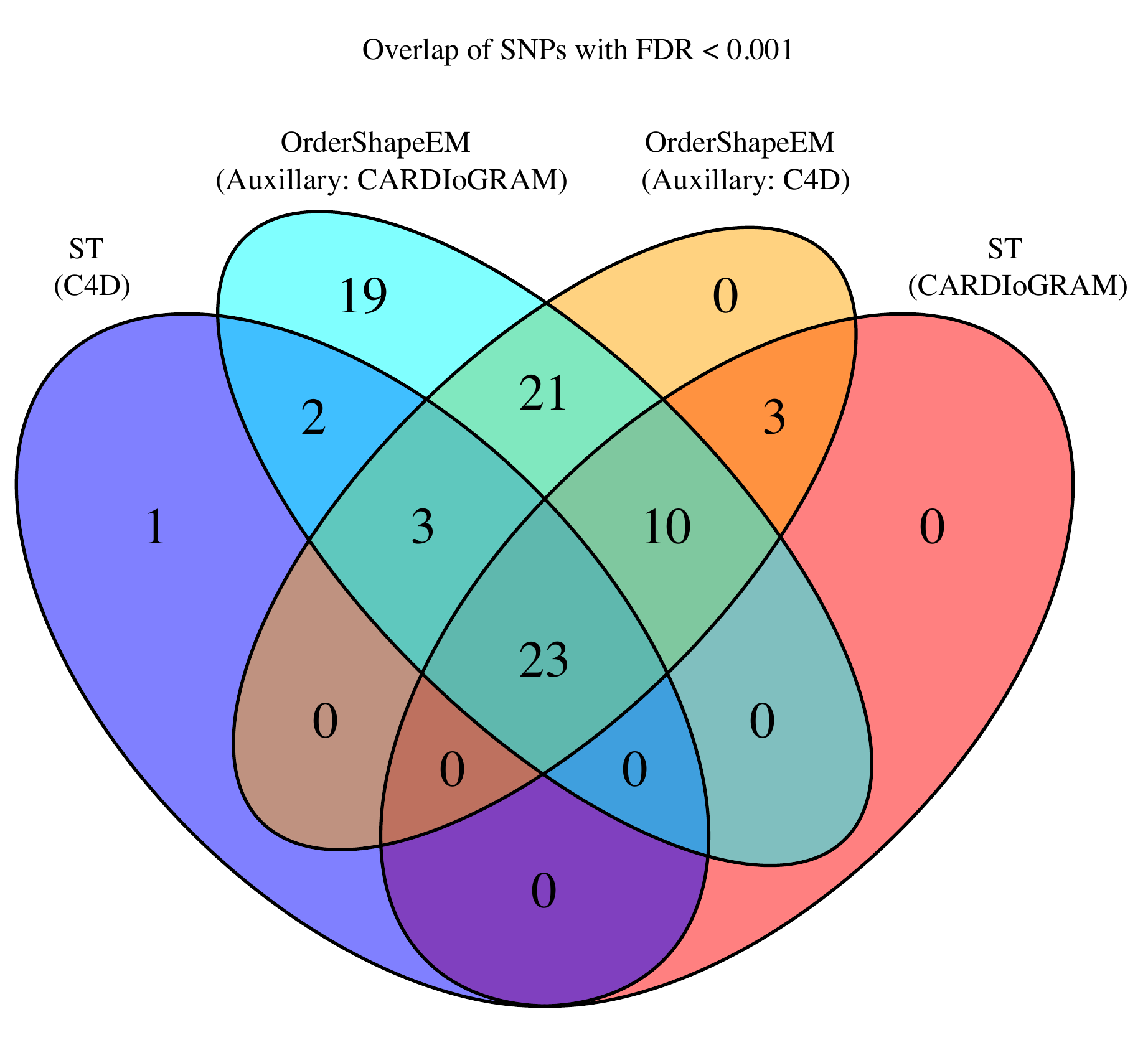}
%\caption[Overlap of associated SNPs with FDR $<$ 0.001.]{ Venn diagram showing the overlap of significant SNPs (FDR $<$ 0.001) between methods using or not using auxiliary information. Left to right: ST procedure on C4D data set; OrderShape EM on C4D data with CARDIoGRAM data as auxiliary; OrderShapeEM on CARDIoGRAM data with  C4D data as auxiliary; and ST procedure on CARDIoGRAM data.} \label{fig:real:4}
\end{figure}

To further examine the identified SNPs based on different methods, Figure \ref{fig:real:4} shows the overlap of significant SNPs via the Venn diagram at FDR level $0.001.$ We observe that there was a significant
overlap of associated SNPs between the two datasets, indicating a shared genetic architecture between the two populations. By using auxiliary information,
OrderShapeEM  recovered almost all the SNPs by ST procedure, in addition to many other SNPs that were missed by the ST procedure. Interestingly, for the $19+21=40$ SNPs that
were identified by OrderShapeEM only,  most of them were located in genes that had been reported being associated with phenotypes or diseases related to the cardiovascular or metabolic
system. It is well known that metabolic disorders such as high blood cholesterol and triglyceride levels are risk factors for CAD.

\section{Summary and discussions}\label{sec:summary}
We have developed a covariate-adjusted multiple testing procedure
based on the Lfdr and shown that the oracle procedure is optimal in the
sense of maximizing the ETP for a given value of mFDR. We
propose an adaptive procedure to estimate the prior probabilities of being null that vary across different hypotheses and the distribution function of the $p$-values under the alternative hypothesis.
Our estimation procedure is built on the isotonic regression which is tuning parameter free and computationally fast.
We prove that the proposed method provides asymptotic FDR control when relevant consistent estimates are available. We obtain some consistency results for the estimates of the prior probabilities of being null and the alternative density under shape restrictions. In finite samples, the proposed method outperforms several existing approaches that
exploit auxiliary information to boost power in multiple testing. The gain in efficiency of the
proposed procedure is due to the fact that we incorporate both the auxiliary
information and the information across $p$-values in an optimal way.

{\color{red}}{
Our method has a competitive edge over competing methods when the signal is weak while the auxiliary information is moderate/strong, a practically important setting where power improvement is critical and possible with the availability of informative prior.  However,  when the auxiliary information is weak , our procedure could be less powerful than the BH/ST procedure. The power loss is more severe under strong and sparse signals.  To remedy the power loss under these unfavorable conditions, we recommend testing the informativeness of the prior order information before the application of our method using, for example, the testing method from \citep{Huang2020}.   We could also  examine the $\hat{\pi}_0$ plot after running our algorithm.  If $\hat{\pi}_0$'s lack variability, which indicates the auxiliary information is very weak,  our method could be less powerful than BH/ST and we advise against using it.
}

Our method is also robust across settings with a very moderate FDR inflation under small feature sizes. However, there are some special cases where our approach does not work well due to the violation of assumptions. In the varying alternative scenario, as suggested by one of the reviewers, we did observe some FDR inflation.  We found this only happens when the order information has inconsistent effects on the $\pi_0$ and $f_1$ (i.e., the more likely the alternative hypothesis, the smaller the effect size). We did not find any FDR inflation if the order information has consistent effects (i.e., the more likely the alternative hypothesis, the larger the effect size).  We believe such inconsistent effects may be uncommon in practice. In the varying null scenario, we observed a severe deterioration of the power of our method and  it has virtually no power when the signal is sparse. This is somewhat expected since our approach assumes a uniformly distributed null p-value.  Therefore, we should examine the p-value distribution before applying our method. We advise against using our method if we see a substantial deviation from the uniform assumption based on the right half of the p-value distribution.

There are several future research directions. For example, it is desirable to extend our method to incorporate other forms of structural information such as group structure, spatial structure or tree/hierarchical structure. %Second, we assume the alternative density belongs to a class of bounded density which is not entirely satisfying in some applications. It is of interest to relax the boundedness assumption. However, it requires nontrivial modifications of existing techniques  (see, e.g., Theorem 7.12 of \cite{van2000}) such as entropy calculation for a particular class of bivariate functions, and we decide to leave it to future research.
Also, the proposed method is
marginal based and it may no longer be optimal in the presence of correlations. We leave these
interesting topics for future research.

\section*{Acknowledgements} The authors would like to thank the Associate Editor
and the reviewers for their constructive comments and helpful suggestions, which substantially
improved the paper. Data on coronary artery disease/myocardial infarction have been contributed by CARDIoGRAMplusC4D investigators and have been downloaded from \url{www.cardiogramplusc4d.org}.
Cao acknowledges partial support from NIH 2UL1TR001427-5, Zhang acknowledges partial
support from NSF DMS-1830392 and NSF DMS-1811747 and Chen acknowledges support from Mayo Clinic Center for
Individualized Medicine.

\bibliographystyle{alpha}

\end{document}